\documentclass[twocolumn]{aastex62}

\usepackage{graphicx}   
\usepackage{amsmath}    
\usepackage{amssymb}    
\usepackage{comment}
\usepackage{xspace}
\usepackage{lastpage}

\usepackage{eso-pic}

\usepackage{comment}

\def\reff@jnl#1{{\rm#1\/}}
\def\aj{\reff@jnl{AJ}}         
\def\araa{\reff@jnl{ARA\&A}}      
\def\apj{\reff@jnl{ApJ}}        
\def\apjl{\reff@jnl{ApJ}}        
\def\apjs{\reff@jnl{ApJS}}       
\def\aap{\reff@jnl{A\&A}}        
\def\aapr{\reff@jnl{A\&A~Rev.}}     
\def\aaps{\reff@jnl{A\&AS}}       
\def\mnras{\reff@jnl{MNRAS}}      
\def\physrep{\reff@jnl{Physics Reports}}
\def\prd{\reff@jnl{Phys.Rev.D}}     
\def\prl{\reff@jnl{Phys.Rev.Lett}}   
\def\pasp{\reff@jnl{PASP}}       
\def\pasj{\reff@jnl{PASJ}}       
\def\nat{\reff@jnl{Nature}}       
\def\jcap{\reff@jnl{JCAP}}   
\def\memsai{\reff@jnl{MemSAI}} 
\def\na{\reff@jnl{New Astronomy}}       
\def\procspie{\reff@jnl{SPIE}}       
\def\pasa{\reff@jnl{PASA}}

\begin{document}

\title{Probing Oort clouds around Milky Way stars with CMB surveys}

\author{Eric J. Baxter}
\affiliation{Department of Physics and Astronomy, University of Pennsylvania, Philadelphia, PA 19104, USA}\email{ebax@sas.upenn.edu}

\author{Cullen H. Blake}
\affiliation{Department of Physics and Astronomy, University of Pennsylvania, Philadelphia, PA 19104, USA}

\author{Bhuvnesh Jain}
\affiliation{Department of Physics and Astronomy, University of Pennsylvania, Philadelphia, PA 19104, USA}

\begin{abstract}
Long-period comets observed in our solar system are believed to originate from the Oort cloud, which is estimated to extend from roughly a few thousand to $10^5$ AU from the Sun.  Despite many theoretical arguments for its existence, no direct observations of the cloud have been reported.  Here, we explore the possibility of measuring Oort clouds around other stars through their emission at submillimeter wavelengths.  Observations with the 545 and 857 GHz bands of the {\it Planck} satellite are well matched to the expected temperatures of Oort cloud bodies (on the order of 10~K).  By correlating the {\it Planck} maps with catalogs of stars observed by the {\it Gaia} mission, we are able to constrain interesting regions of the exo-Oort cloud parameter space, placing limits on the total mass and the minimum size of grains in the cloud.  We compare our measurements with known debris disk systems -- in the case of Vega and Fomalhaut we find a significant excess that is in agreement with measurements from {\it Herschel}.  We use the measurements around Fomalhaut to constrain a possible exo-Oort cloud of that system.   We explore an observed excess around the brightest and nearest stars in our sample as arising from possible exo-Oort clouds or other extended sources of thermal emission.  We argue that future CMB surveys and targeted observations with far-infrared and millimeter wavelength telescopes have the potential to detect exo-Oort clouds or other extended sources of thermal emission beyond $\sim 1000$ AU from the parent stars.
\vspace{1cm}
\end{abstract}

\section{Introduction}
The observation of comets passing through the inner solar system led \cite{Oort:1950} to hypothesize the existence of a spherical cloud of distant icy bodies, now known as the Oort cloud (OC).  Since then, a number of additional theoretical arguments in support of Oort's hypothesis, as well as a more detailed understanding of the cloud's expected properties, have emerged \citep[for a review, see e.g.][]{Dones:2004, Dones:2015}.  However, to date, no direct observation of the Sun's OC has been made.

The Oort cloud, sometimes also called the \"Opik-Oort cloud, is believed to originate from a population of small, icy bodies within 50 AU of the sun that were present in the young solar system. Orbital perturbations caused by the giant planets would, in a short time, increase the orbital energies of many of these bodies onto highly elliptical orbits with very large semi-major axes. Bodies in regions of relative dynamic stability could remain, mainly in the ecliptic plane, resulting in the Kuiper Belt and ecliptic comet populations that we see today. For bodies with eccentric orbits and semi-major axes of tens of thousands of AU, interactions with nearby stars, the galactic potential, and nearby molecular clouds can stabilize these orbits by increasing the perihelion distances of the orbits so that perturbations by planets in our own solar system are no longer dynamically important. Assuming these gravitational interactions  are isotropic,  the expected result is a population of comets having semi-major axes between a few thousand and tens of thousands of AU and inclinations that are randomly distributed relative to the ecliptic plane.  Since other stars likely experienced similar histories, it is reasonable to expect that they may also host their own Oort clouds, which we refer to as exo-Oort clouds (EXOCs). In fact, there have been several reported detections of exo-comets in the literature through transits (e.g.~\citealt{Rappaport:2018}) as well as the spectral signatures of evaporating small, icy bodies (e.g.~\citealt{Welsh:2016}). Other authors have investigated the fate of EXOCs as stars evolve and proposed potentially detectable signatures associated stellar remnants (e.g. \citealt{Stone:2015}).

Directly detecting the OC orbiting the sun, or the EXOC of another star, is extremely challenging. The vast distances of these bodies from their parent stars mean that they are faint in reflected light. While there may be a very large number of OC bodies, their total surface area results in an effective OC optical depth that is extremely small, $\tau<10^{-6}$, even if the OC is very massive (above 100$M_\earth$) and contains a large number of small (micron-sized) particles. Stellar occultations provide a promising avenue for directly detecting bodies in our OC (see \citealt{Lehner:2016, Ofek:2010}), but the events are very short (less than 1~second in duration) and rare, meaning that a large number of stars have to be observed at very high cadence. In principle, the thermal emission from the OC imprints a distortion of the black body spectrum of the CMB, but even for optimistic assumptions about the mass of our OC, this signal is too small to be observed with existing CMB experiments (\citealt{Babich:2007, Babich:2009,Ichikawa:2011}).  \citet{Cowan:2016} explored the possibility of detecting thermal emission from the hypothetical `Planet Nine,' believed to reside in the Oort cloud, using existing and future CMB experiments, and found that detection prospects were promising.  Detecting thermal emission from OCs around nearby, bright stars is another possibility. Particularly at long wavelengths close to the peak of the EXOC blackbody, the large aggregate surface area of emitters in the EXOC means that thermal emission from the central star itself is expected to be orders of magnitude smaller than that from the EXOC, even though the star is much hotter than the typical EXOC body. \citet{Stern:1991} used IRAS data to search for excess mid-IR emission in the vicinities of a small number of nearby, bright stars in an attempt to place limits on the total masses of their EXOCs. We will pursue longer wavelengths that are more suitable for EXOC detection using a much larger sample of stars. 

Given that the OCs of nearby stars may be tens of thousands of AU in diameter, they may subtend tens of arcminutes on the sky as seen from Earth. This makes it possible to use high spatial resolution, wide-area CMB surveys like \textit{Planck} (\citealt{Planck:2018}) to place limits on the average excess thermal emission at millimeter and submillimeter wavelengths using large samples of nearby stars having precise distances now measured by \textit{Gaia} (\citealt{Gaia:2018}).  Such emission can be distinguished from more localized debris disk or point source emission using the large difference in scales between such emission and that of an EXOC.  Typical debris disks extend to only a few hundred AU at most, orders of magnitude smaller than the expected sizes of EXOCs. 

In this work, we analyze \textit{Planck} 545 and 857 GHz maps to place  limits on the total non-stellar thermal emission within the EXOC regions of a sample of main sequence stars within 300~pc of the Sun identified using {\it Gaia} data.  We use stacked measurements across these stars to place limits on the properties of EXOCs.  Since it is  not known how generic EXOCs are, we also explore for the brightest stars the possibility that only some fraction of them have EXOCs.  We argue that current and future CMB surveys may offer the possibility of improved limits on EXOC properties, and explore the possibilities for future detections using targeted observations.  

The paper is organized as follows: in \S\ref{sec:model} we describe the expected properties Oort clouds, and our model for their thermal emission; in \S\ref{sec:data} we describe the datasets used in this analysis; in \S\ref{sec:measurements} we describe our procedure for measuring the thermal emission signal around the {\it Gaia} stars.  Our results are presented in two parts: in \S\ref{sec:stacked_results} we show the results averaged over many of the stars in the sample, while in \S\ref{sec:hot_results}, we investigate possible signals around the closest and hottest stars.  We conclude and discuss prospects for future measurements in \S\ref{sec:discussion}.

\section{Modeling}
\label{sec:model}

We are interested in modeling the thermal emission signal from EXOC bodies.  The expected signal is sensitive to several aspects of the Oort cloud, including its total mass, the size distribution of the Oort cloud bodies, and the intensity of radiation from its parent star. As we will show, for reasonable assumptions, the expected temperatures of Oort cloud objects are tens of Kelvin.  For these temperatures, the peak emission is at submillimeter wavelengths, so the natural datasets to pursue the signal are far-infrared surveys or surveys designed to map the Cosmic Microwave Background (CMB) at millimeter wavelengths.  

The {\it Planck} satellite mapped the CMB in nine  frequency channels and is well suited to the pursuit of EXOC signals. We will focus on the highest two frequency channels, centered at 545 and 857 GHz, for the remainder of this paper and return to a discussion of datasets in other frequencies in the Discussion section. Since the wavelength corresponding to 857 GHz is 0.35~mm, it is close to the peak emission of EXOC bodies at a temperate of $\sim 10K$. We describe both the {\it Planck} data and the {\it Gaia} star catalog with which we correlate it in more detail in \S\ref{sec:data}. 

\subsection{Expected properties of Exo-Oort clouds}

The total mass of our Oort cloud is estimated from the rate of long-period comets observed to pass through the inner solar system. Oort himself estimated an OC comet flux of one per year within the inner 1.5~AU of the solar system, leading him to estimate that the total number of OC comets is approximately $10^{11}$ with a total mass of the OC of approximately 0.3~$M_\earth$. Current estimates of the OC mass, made using updated comet flux data and assuming a comet size distribution, range up to 10~$M_\earth$ (see \citealt{Dones:2004} and references therein). This value is highly uncertain, in part because it is difficult to account for inner OC bodies that seldom venture into the inner solar system for us to observe. Given our current observational understanding of the OC, total masses of up to 100$M_\earth$ are not obviously ruled out \citep{Bauer:2017}.
Our Oort cloud is believed to extend roughly from a few thousand to $10^5$ AU, though there are significant uncertainties in demarcating its extent.  At distances of $10^{5}$ AU or more, bodies become unbound to the sun and may be lost to interstellar space.  One might expect the size of the EXOC to be related to the Hill Sphere of the central star, in which case the size of the Oort cloud would increase with stellar mass as $r_{\rm max} \propto M_{*}^{1/3}$. However the age of the star, its planetary architecture and possible membership in a star cluster can have a significant impact on the extent and mass of EXOCs. 

Based on direct observations of long-period comets as they move through the inner solar system, and the recent discoveries of possible inner-OC bodies like Sedna (\citealt{Brown:2004}), we expect that the OC bodies are primarily icy in composition. Recently, it has also been suggested that there is a small but significant population of scattered asteroids with rocky compositions in the OC (\citealt{Shannon:2015}). Since small, icy OC bodies are thought to be generated from the disintegration of larger ones, the internal structural properties of the OC objects play an important role in determining their overall size distribution \citep{Pan:2005}. Currently, we have only a rough understanding of the bulk properties of outer solar system bodies. Measuring the masses of these objects is only possible if they have measured radii and small moons with measured orbital periods. Radiometric measurement of the objects' radii is challenging (\citealt{Brown:2017}), making their volumes highly uncertain even if the masses are well measured using moons. The densities of only a handful of outer solar system bodies are directly measured, but Pluto's density of $1.9$~g~cm$^{-3}$ is likely indicative of the densities of OC bodies. 

Inferring the overall size distribution of the bodies in the OC is a major observational challenge. Observations of long-period comets and outer solar system bodies like Sedna are highly biased toward larger, closer, and therefore brighter, bodies. Theoretically, it is expected that the overall size distribution of OC bodies should follow a power-law. \cite{Pan:2005} simulated the evolution of a population of icy bodies with very little internal strength and found that  there should be a break in the overall size distribution of OC bodies at radii of approximately 40~km, which decreases the expected number of very large bodies. This is broadly consistent with the observational results of \cite{Bernstein:2004}, who used \textit{HST} to carry out a pencil-beam survey for Kuiper Belt bodies, which have much smaller semi-major axes than the OC bodies we consider here. 

There is also considerable uncertainty in the small-scale distribution of Oort cloud bodies.  Collisions between comets can produce small grains and dust down to scales of about 1~$\mu$m.  However, several effects tend to drive the smallest dust grains out of the solar system.  These include radiation pressure from the central star, stripping by the ISM, and stellar winds.  At reasonable Oort cloud distances, though, the effects of the stellar wind are typically negligible.  \citet{Howe:2014} argue that grains smaller than about 10~$\mu$m are stripped from the solar system by the passage of the solar system through the ISM.  However, smaller grains that have been recently produced in collisions may still be present.  No direct measurement of the size distribution of bodies in the OC has been made, but in the future observations of background stars occulted by bodies in our OC may be able to directly constrain the overall size distribution  down to about 1~km in radius (\citealt{Lehner:2016}). 

The temperature of Oort cloud bodies is set by equilibrium between absorption of light from the central star, interstellar light, and the cosmic background radiation, and thermal emission from the objects.  In principle, radioactive decays could also provide a heat source, but this is expected to be subdominant to radiative heating for the small objects that dominant the EXOC thermal emission.  The material properties and sizes of the bodies determines how efficiently they absorb and emit radiation, and therefore their equilibrium temperatures.  For the OC grains of interest, the thermal emissivities at relevant wavelengths are expected by be less than one, leading to temperatures on the order of tens of Kelvin (see details of the dependence on grain size below).  For stars more luminous than the Sun, these temperatures can be significantly higher for a given orbital distance. In principle, Oort cloud objects could also reflect stellar light, resulting in a different equilibrium temperature.  The relatively high geometric albedo of Sedna (A$=0.32\pm0.06$;~\citealt{Pal:2012}) confirms the predominantly icy surface composition of that body.  However, other Kuiper belt objects appear to have much lower albedos (\citealt{Brown:2017}), and we will assume low albedo in the EXOC model developed below.

\subsection{Exo-Oort cloud model}

Given the above discussion, we now develop a model for the EXOC thermal emission signal.  We assume that the Oort cloud mass, $M$, is proportional to a power law function of the mass of its parent star, $M_*$:
\begin{eqnarray}
\label{eq:oort_mass}
M = A_{M} (M_{*}/M_{\odot})^{\mu}.
\end{eqnarray}
We set $\mu = 1$, but find that our results are fairly insensitive to the precise value of $\mu$ given the narrow range of stellar masses considered.  We treat the normalization of this relation, $A_{M}$, as a parameter of our analysis.  As noted above, the total mass of our own OC is poorly constrained, but reasonable estimates range from about 0.3 $M_{\oplus}$ to $20 M_{\oplus}$.

We use $r$ to represent the distance between a point in the EXOC and its central star, and  $R$ to represent the projection of that distance onto our line of sight towards the star (i.e. the impact parameter).
We model the radial dependence of the mass density of the Oort cloud, $\rho(r)$, with a power law (\citealt{Duncan:1987, Howe:2014}) 
\begin{eqnarray}
\label{eq:rho}
\rho(r) = \begin{cases}
 A_{\rho} (r/r_{\rm min})^{-\gamma} &\text{if $r_{\rm min} < r < r_{\rm max}$}\\
0 &\text{otherwise},
\end{cases}
\end{eqnarray}
where the normalization, $A_{\rho}$, is set such that the total mass of the cloud is $M$:
\begin{eqnarray}
A_{\rho} = \frac{M (3-\gamma)}{4\pi \left(r_{\rm max}^{3}(r_{\rm max}/r_{\rm min})^{-\gamma} - r_{\rm min}^{3} \right)}.
\end{eqnarray}
We assume that Oort cloud bodies have constant density of $\rho_0$; we choose $\rho_0 = 1\,{\rm g}~{\rm cm^{-3}}$ as a fiducial value.  Following \citet{Howe:2014}, we adopt a fiducial value of $\gamma = 3.5$.  Given the expected weak scaling of $r_{\rm max}$ with stellar mass ($r_{\rm max} \sim M_{*}^{1/3}$ if $r_{\rm max}$ is determined by the star's Hill sphere) and the large uncertainties associated with the radius of our own OC, we ignore possible scaling of $r_{\rm max}$ with stellar mass and simply treat $r_{\rm max}$ as a free parameter below.

We model the probability distribution of the radii of Oort cloud bodies, $a$, as a broken power law that is independent of distance from the star:
\begin{eqnarray}
\label{eq:size_distribution_small}
P(a) = \begin{cases}
 A_P (a/a_{\rm break})^{-\beta_1} &\text{if $a_{\rm min} < a < a_{\rm break}$}\\
A_P (a/a_{\rm break})^{-\beta_2}  &\text{if $a_{\rm max} > a > a_{\rm break}$}, \\
0 &\text{otherwise}
\end{cases}
\end{eqnarray}
where $a$ is the radius of the Oort object and $A_P$ is a normalization factor.  The small end power law is  expected to be $\beta_1 \sim 3.7$ and the large end power law is expected to be $\beta_2 \sim 5$ \citep{Pan:2005}.  We explore  several choices of $a_{\rm min}$ ranging from $1 \mu{\rm m}$ to $50 \mu{\rm m}$.  Following \citet{Pan:2005}, we set $a_{\rm break} = 40\,{\rm km}$.

The geometric absorption coefficient at distance $r$ from the central star is calculated by integrating over the particle distribution:
\begin{eqnarray}
\alpha_{\nu}(r) &=&  \int_{a_{\rm min}}^{a_{\rm max}}  \, \pi a^2 \frac{\rho(r)}{\int_{a_{\rm min}}^{a_{\rm max}} (4\pi/3)  (a')^3 \rho_0 P(a')da' } P(a)  da. \nonumber \\
&=& \frac{3\rho(r)}{4\rho_0 a_{\rm break}} \left[ \frac{  \frac{1}{3-\beta_1} - \frac{1}{3 - \beta_2} - \frac{u^{3-\beta_1}}{3-\beta_1} }{ \frac{1}{4-\beta_1} - \frac{1}{4-\beta_2} } \right],
\end{eqnarray}
where $u = a_{\rm min}/a_{\rm break}$, and in the second line we have assumed that $a_{\rm min} \ll a_{\rm break}$, $a_{\rm max} \gg a_{\rm break}$, $\beta_1 < 4$ and $\beta_2 > 4$.  The total optical depth of the cloud for a line of sight with impact parameter $R$ is then
\begin{eqnarray}
\tau(R) = \int_{-\infty}^{\infty} a_{\nu}\left( \sqrt{R^2 + s^2} \right) ds.
\end{eqnarray}

\subsection{Temperature profile}

Because of their large cross-sectional area per mass, the small grains will dominate the thermal emission from the EXOCs.  For a grain temperature of 10 K, the wavelength at peak emission given by Wien's law is roughly 300 $\mu$m, whereas we are assuming that the smallest OC grains have radius $a\sim10\mu$m.  Consequently, the relevant wavelengths of light for thermal emission are long compared to the sizes of the smallest grains in our model.  Assuming the grains have a dirty ice composition, we therefore expect the emissivities of the grains to be inversely proportional to wavelength, i.e. $\epsilon(\lambda) = \lambda_0/\lambda$ (\citealt{Dwek:1980}).  In thermodynamic equilibrium, we then have
\begin{multline}
T(r) = 5.2 K \left[(1 K)^{-4}(1-A)\left( \frac{\lambda_0}{1\mu m}\right)^{-1} \right.\\ 
\left. \left( T_{\rm bg}^4 + \frac{L_{*}}{16 \pi \sigma_{sb} r^2} \right) \right]^{1/5},
\end{multline}
where $A$ is the albedo, and we set the background temperature $T_{\rm bg} = 3.5\,K$ \citep{Stern:1991}.  We set $A = 0.03$ following \citet{Stern:1991} and adopt $\lambda_0 = 10\,\mu$m as a reasonable choice for dirty ice grains of this size \citep{Morales:2016}.  To model the stellar emission, we  relate the observed temperatures of the {\it Gaia} stars to their luminosities and masses by interpolating between the locus points from \citet{Eker:2018}.

\subsection{Signal profile}

We assume that the Oort cloud bodies emit with a Planck blackbody source function, modulated by their emissivities.  The specific intensity from the EXOC at projected radius $R$ and frequency $\nu$ is then  
\begin{multline}
I_{\nu}(R) = \int_{-\infty}^{\infty}  \,\epsilon(c/\nu)\alpha_{\nu}\left( \sqrt{R^2 + s^2} \right) \\
B_{\nu}\left( T \left(\sqrt{R^2 + s^2} \right) \right) ds,
\end{multline}
where $B_{\nu}(T)$ is the Planck function.  

The signal in the {\it Planck} 857 GHz band is found by integrating $I_{\nu}$ over the bandpass of that channel:
\begin{equation}
I(R) = \int \, I_{\nu}(R) B_{857}(\nu) d\nu,
\end{equation}
where $B_{857}(\nu)$ is the bandpass function.  We approximate this function as a top-hat with amplitude 0.8 over the frequency range from 750 to  1000 GHz. 

Finally, we must account for the impact of the {\it Planck} beam.  We approximate the {\it Planck} beam with a 2D Gaussian, $F(\vec{\theta}) \propto \exp(-\theta^2/(2\sigma_{\rm beam}^2))$, with $\sigma_{\rm beam} = \theta_{\rm FWHM}/(\sqrt{8 \ln 2})$, where $\vec{\theta}$ is the angular separation vector between the parent star and a point of interest. The specific intensity after the application of the beam, $I^{\rm beam}(\vec{\theta})$, is given by
\begin{eqnarray}
I^{\rm beam}(\vec{\theta}) \propto \int I(\vec{\theta}') F(\vec{\theta}' - \vec{\theta})  d^2\theta',
\end{eqnarray}
where the constant of proportionality is fixed by the requirement that $\int I^{\rm beam}(\vec{\theta}) d^2\theta = \int I(\vec{\theta}) d^2\theta$.  While the \textit{Planck} beam does have some far-wing sensitivity, \citep{Planck:HFI}, it is too low to be important for this analysis. 

\begin{figure}\includegraphics[scale=0.45]{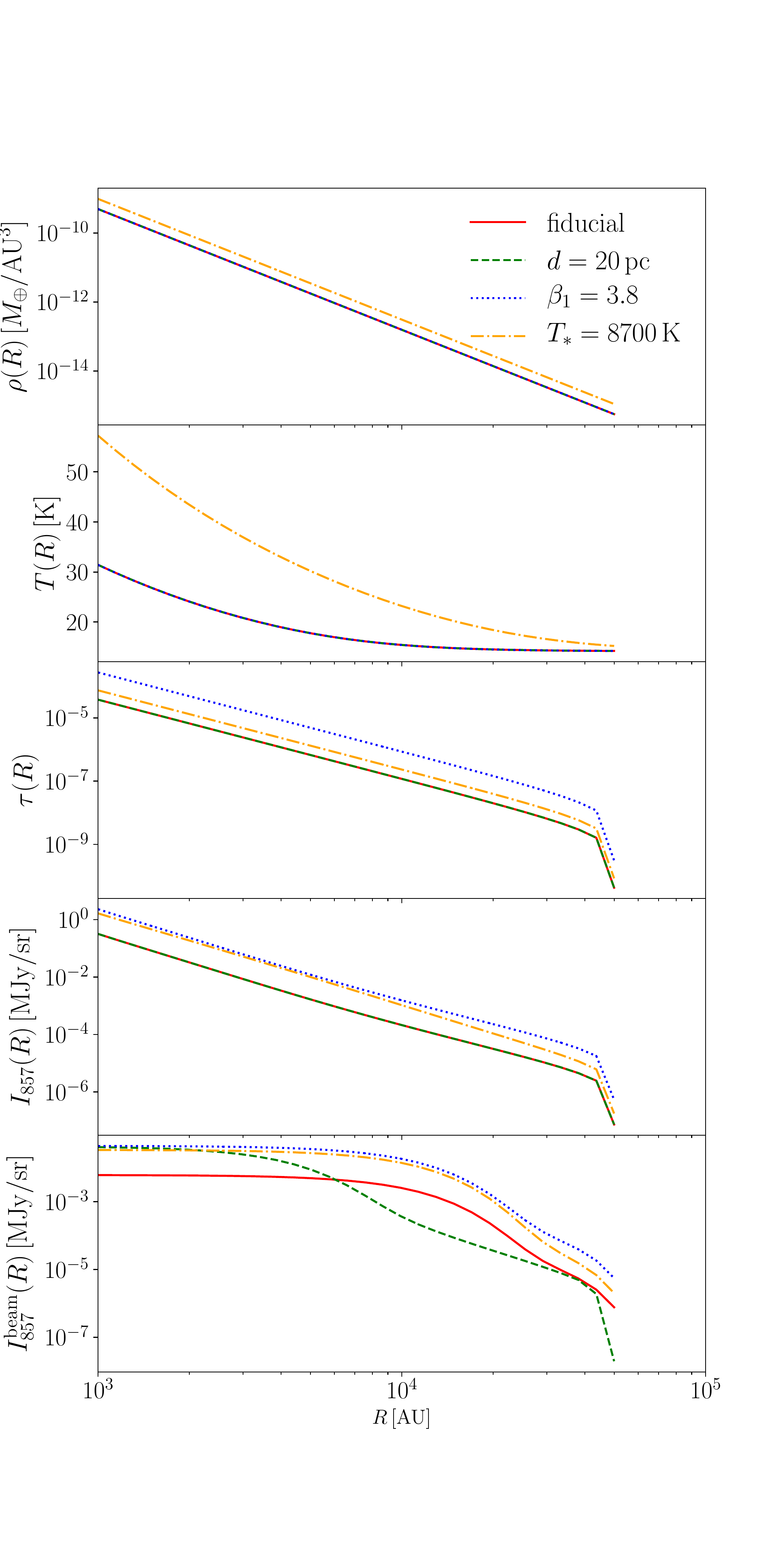}
\caption{\label{fig:model_curves}
 Oort cloud properties as a function of projected radial distance from the central star, $R$.  The fiducial model assumes $A_M = 5 M_{\oplus}$, $r_{\rm max} = 5\times 10^4\,{\rm AU}$, $a_{\rm min} = 10^{-6}\,{\rm m}$, $\beta_1 = 3.7$, $\beta_2 = 5$, $a_{\rm break} = 40 \,{\rm km}$, $\gamma= 3.5$, $\rho_0 = 1 \, {\rm g}/{\rm cm}^3$, and $d = 60\,{\rm pc}$.}
\end{figure}

We show the expected density, temperature, optical depth and 857 GHz EXOC profiles (with and without the beam) for several parameter choices in Fig.~\ref{fig:model_curves}.

\subsection{Additional sources of submillimeter emission around stars}

In addition to possible Oort clouds, many stars are known to host debris disks which can be strong emitters at submillimeter wavelengths.  Debris disks are typically understood to be a post-planet formation phase of protoplanetary disks. Many debris disks around nearby stars have been detected via their thermal emission by IRAS, Herschel and ALMA, as well as in scattered light by HST (see review by \citealt{Hughes:2018}).  The presence of debris disks presents a challenge for Oort cloud detection, since the total emission from these disks may dominate that of an Oort cloud by an order of magnitude or more.  However, physical differences between debris disks and Oort clouds should make separating them possible.  In particular, debris disks are expected to be confined to a plane, while Oort clouds are expected to have spherical geometry.  Additionally, the physical extent of a debris disk is of order 100 AU (\citealt{Hughes:2018}), while Oort clouds could extend out to $10^5$ AU.  A high resolution experiment could separate out the Oort cloud emission using both of these differences. An additional complication is the possible ``halo'' of particles blown out of debris disks due to radiation and stellar winds. We discuss this possibility further below. 

Stars can produce both thermal and non-thermal radio emission through a variety of processes (\citealt{Gudel:2002}), and emission is observed from many different types of stars in the 1 to 10 GHz range. This emission can be related to chromospheric activity and flares, stellar interaction in binary systems, or accretion processes in very young star systems. High energy electrons from a stellar wind may also generate radio emission through gyrosynchrotron processes. 
Relatively little is known about the quiescent radio emission from stars at the much higher frequencies considered here, though millimeter and submillimeter emission from the photospheres of a small number of stars has recently been directly measured (\citealt{White:2018, Anglada:2018}). 

\section{Data}
\label{sec:data}

\subsection{The {\it Gaia} catalog}
We use data from the \textit{Gaia} DR2 release to identify stars within 300~pc of the Sun (\citealt{Gaia:2018}). This data release contains proper motions and parallaxes for more than $10^{9}$ stars down to a limiting magnitude of G=21. The {\it Gaia} satellite also obtains low-resolution spectroscopy, from which broad-band photometry is extracted and stellar parameters are estimated (\citealt{Andrae:2018}). We focus our analysis on nearby stars with $d < 300\,{\rm pc}$. For main sequence stars in this volume, the {\it Gaia} parallax measurements can have errors of less than $\pm$0.1 mas. The {\it Gaia} data set is virtually complete for main sequence stars in this volume. 

\subsection{{\it Planck} data}

The {\it Planck} satellite\footnote{\url{http://www.esa.int/Planck}} observed the full sky in nine bands ranging from 30 GHz to 857 GHz over the course of four years.  In this analysis, we use primarily the 545 and 857 GHz maps constructed from observations by the {\it Planck} High Frequency Instrument \citep{Planck:HFI}.  The 545 and 857 bandpasses of {\it Planck} are well matched to the expected frequency of peak emission for Oort cloud bodies.  The approximate beam sizes for the 545 and 857 GHz bandpasses are 4.8 and 4.6 arcminutes FWHM, respectively \citep{Planck:HFI}.  We use the publicly available maps at \url{https://pla.esac.esa.int/}.  We will also make use of the 217 GHz and 353 GHz maps; for these maps, we convert from $K_{\rm CMB}$ units to intensity units using the conversion factors listed in \citet{Planck:conversion}.

\begin{figure}
\includegraphics[scale=0.45]{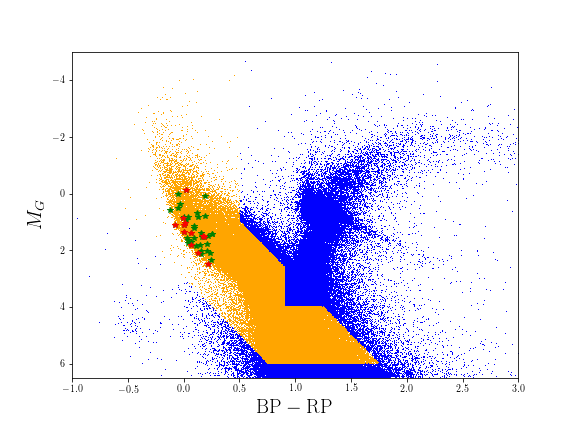}
\caption{\label{fig:star_selection} The selection of {\it Gaia} stars used in this analysis.  Orange points indicate stars that pass the baseline selections of the analysis, intended to identify main sequence stars.  Blue points represent a wider selection to show the regions of color space that are excluded.  The cuts remove stars that are much fainter than the Sun (which may not have Oort clouds) and giant stars (which are cooler and may have emission in the wavebands of interest related to dust produced during the AGB phase).  The star symbols indicate the selection of $T > 8000\,{\rm K}$ stars used in the analysis of \S\ref{sec:hot_results}.  Green stars indicate those that do not have a signal-like excess, while red stars indicate those that do.}
\end{figure}

\begin{figure}
\includegraphics[scale=0.4]{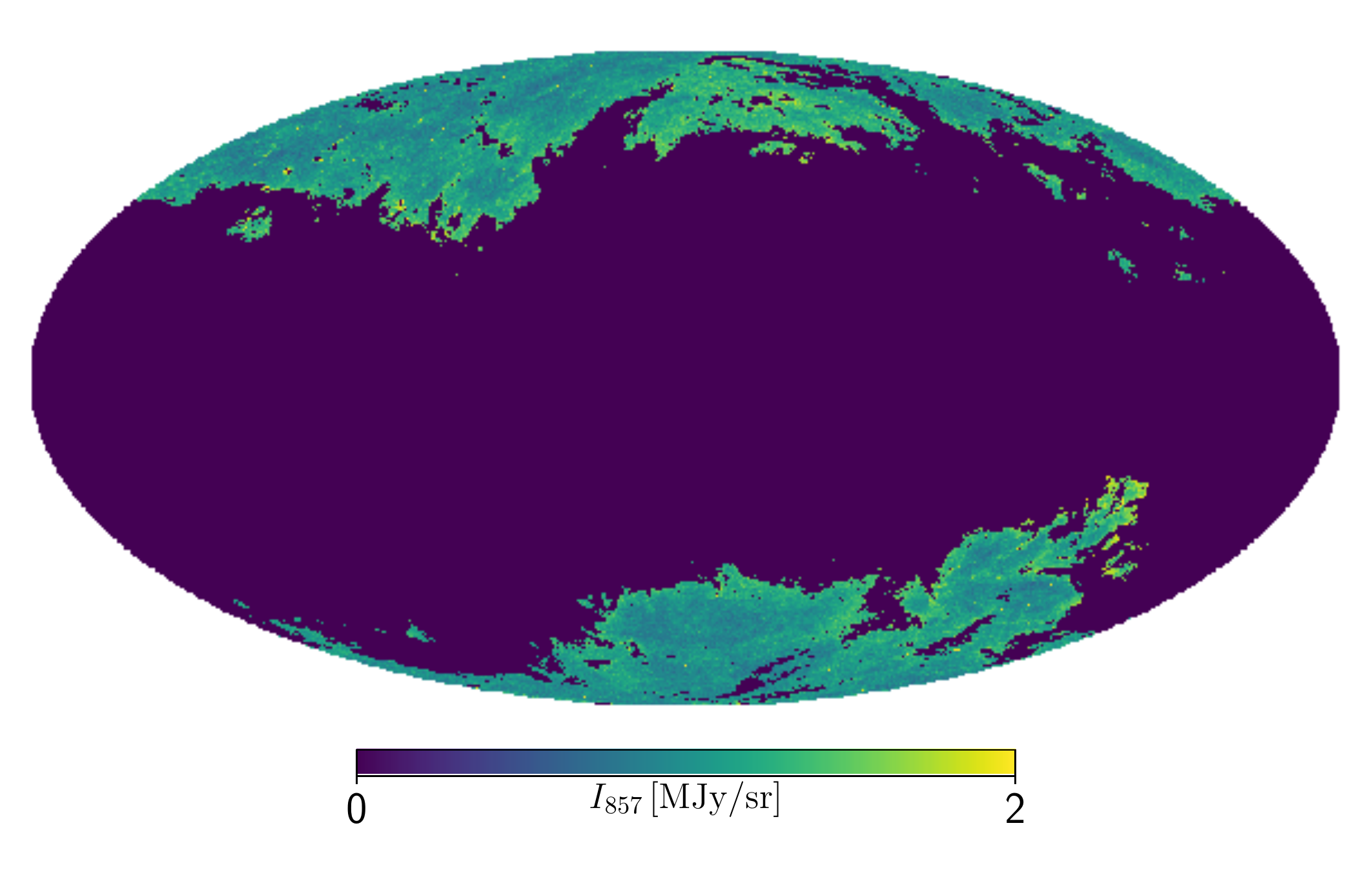}
\caption{\label{fig:mask} The signal at 857 GHz after imposition of the HI mask.  Dark blue regions are masked because they have HI column densities in the top 80\% across the sky.}
\end{figure}

\section{Methodology}
\label{sec:measurements}

We focus our analysis on main sequence stars that have luminosities comparable to or brighter than the sun.  Significantly smaller stars may not posses their own Oort clouds, and any Oort cloud that they do possess will likely be less luminous because of its lower temperature.  Similarly, we remove possible white dwarfs as it is unclear whether such stars would be expected to have EXOCs.  Giant stars, on the other hand, can have significant thermal emission related to ejected gas and dust that could masquerade as an Oort cloud signal (\citealt{Ventura:2018}).  Note, though, that emission from giant stars would not be resolvable by the {\it Planck} beam, whereas Oort cloud emission could be resolved for nearby stars.

We accomplish the selection motivated above with several cuts:
\begin{itemize}
\item The absolute magnitude in the {\it Gaia} G band, $M_G$, is less than 6.0.
\item We remove stars that have $M_G < 4$ and color $BP-RP > 0.9$, since these stars are possible giants. 
\item Stars with $M_G <4\times (BP-RP)-1$ and $(BP-RP) > 0.5$ are also removed as possible giants.
\item stars with $M_G >4\times (BP-RP) + 3$ are removed as possible white dwarfs. 
\end{itemize}
The stellar selection is illustrated with the orange points in Fig.~\ref{fig:star_selection}.

We measure the average temperature of the {\it Planck} 545 and 857 GHz maps in annuli around the selected {\it Gaia} stars.  The annuli are defined in terms of physical projected distance from the star, from $R_{\rm min} = 0\,{\rm AU}$ to $R_{\rm max} = 10^{5}\,{\rm AU}$ in six linearly spaced bins.  We use the stellar distance measurements from \textit{Gaia} to convert these projected radial bins into angular bins for each star.  In \S\ref{sec:hot_results} we will perform measurements around the closest stars in our sample, for which a different choice of radial bins is motivated.

We reduce the impact of non-EXOC galactic emission on our measurements by imposing a conservative sky mask in the analysis.  The mask is derived from a map of  HI column density by \citet{Lenz:2017}, which has been shown to correlate strongly with thermal emission from galactic cirrus at low HI column density.  We mask those pixels that have column densities in the top 80\% of pixels in this map.  Furthermore, if the apertures used to estimate the background (see below) near to a star intersects the mask, that star is not included in the analysis, since presumably it lives in or near a region of high background emission.  The 857 GHz {\it Planck} map with this mask applied is shown in Fig.~\ref{fig:mask}.  Additionally, we use the PSCz catalog from \cite{Saunders:2000} to mask IRAS-detected galaxies with an aperture of 5' and NGC objects with a mask of 15'. 

To further reduce the impact of galactic contamination on the measurements, we subtract an estimate of the local background flux for each star.  This estimate of the background is derived by averaging  the {\it Planck} map in an annulus of width 10' centered on the star.  The angular size of the inner radius of the annulus is either (a) the angular scale corresponding to $10^{5}\,{\rm AU}$ or (b) 10'; we choose whichever option yields the larger aperture size.  To prevent bright patches from biasing the background estimate in the apertures, when computing the average in the annulus, we exclude pixels whose flux is in the bottom or top 10\% of that aperture.   

Finally, to remove possible spurious signal from the coupling of background fluctuations to the mask, when performing stacked measurements we subtract from these measurements the signal measured around random points uniformly distributed across the mask.  This typically results in a small change to the measured signal.

The above analysis choices are well motivated, but of course are not unique.  One could imagine, for instance, choosing different masking thresholds, different aperture sizes, or different procedures for estimating the background flux near the stars.  We discuss the impact of varying the fiducial analysis choices in \S\ref{sec:variations_stacking}.

\section{Constraints on EXOC emission from stacking analysis}
\label{sec:stacked_results}

We first average the intensity measurements around the {\it Gaia} stars in bins of $M_G$ and distance from Earth.  Averaging the signals from many stars allows us to beat down instrumental and background noise in the {\it Planck} maps.  Binning in absolute magnitude reduces the possibility that the (much more numerous) faint stars might cause an Oort signal around brighter stars to be averaged down.  Binning in distance is useful since the apparent sizes of EXOCs relative to the {\it Planck} beam and relative to scales of variation in galactic backgrounds will change as the distance to the stars changes.

We note, though, that averaging the signals from many stars may be disadvantageous if not every star hosts an EXOC.   The proposed mechanism of formation of our own Oort cloud relies on the presence of the giant planets to perturb the orbits of solar system objects.  If a star does not have any giant planets, it is possible that it might not form an EXOC.  Including such stars in the averaging process could reduce the signal averaged across many stars.  We explore an alternative to averaging across many stars in \S\ref{sec:hot_results}.

For a resolved EXOC, the surface brightness will be independent of distance, $d$, but the number of resolution elements across the cloud will go down as $d^{-2}$.  Assuming constant stellar density in our local neighborhood, the number of stars in a shell of thickness $\Delta d$ at distance $d$ is proportional to $d^2 \Delta d$, so the signal-to-noise for each radial shell of unit thickness is constant after averaging over resolution elements (this same expectation also holds true for larger distances where the EXOC may be unresolved, although the number of resolution elements will remain fixed in that case).   However, the scale height of the Milky Way stellar disk is roughly 300~pc; consequently, beyond 300~pc, the number of stars per radial shell will only grow as $d \Delta d$. Hence we set 300~pc as the maximum distance used in our analysis. 

When binning the stars in distance, we prefer to not mix vastly different scales of galactic emission; this motivates the use of roughly logarithmic bins in $d$ (which will lead to the higher distance bins having more signal-to-noise).  Based on these considerations, we use three distance bins between 80 and 300 pc: $[80,120]$, $[120,180]$, and $[180,300]$.  We will explore the stars with $d < 80~{\rm pc}$ in \S\ref{sec:hot_results}.

We also restrict the analysis in this section to stars with $2 < M_G < 6$ (using two bins).  This restriction ensures that we do not use stars significantly fainter than the sun.  We will explore measurements around the brightest stars in \S\ref{sec:hot_results}.

\begin{figure*}
\centering
\includegraphics[scale=0.6]{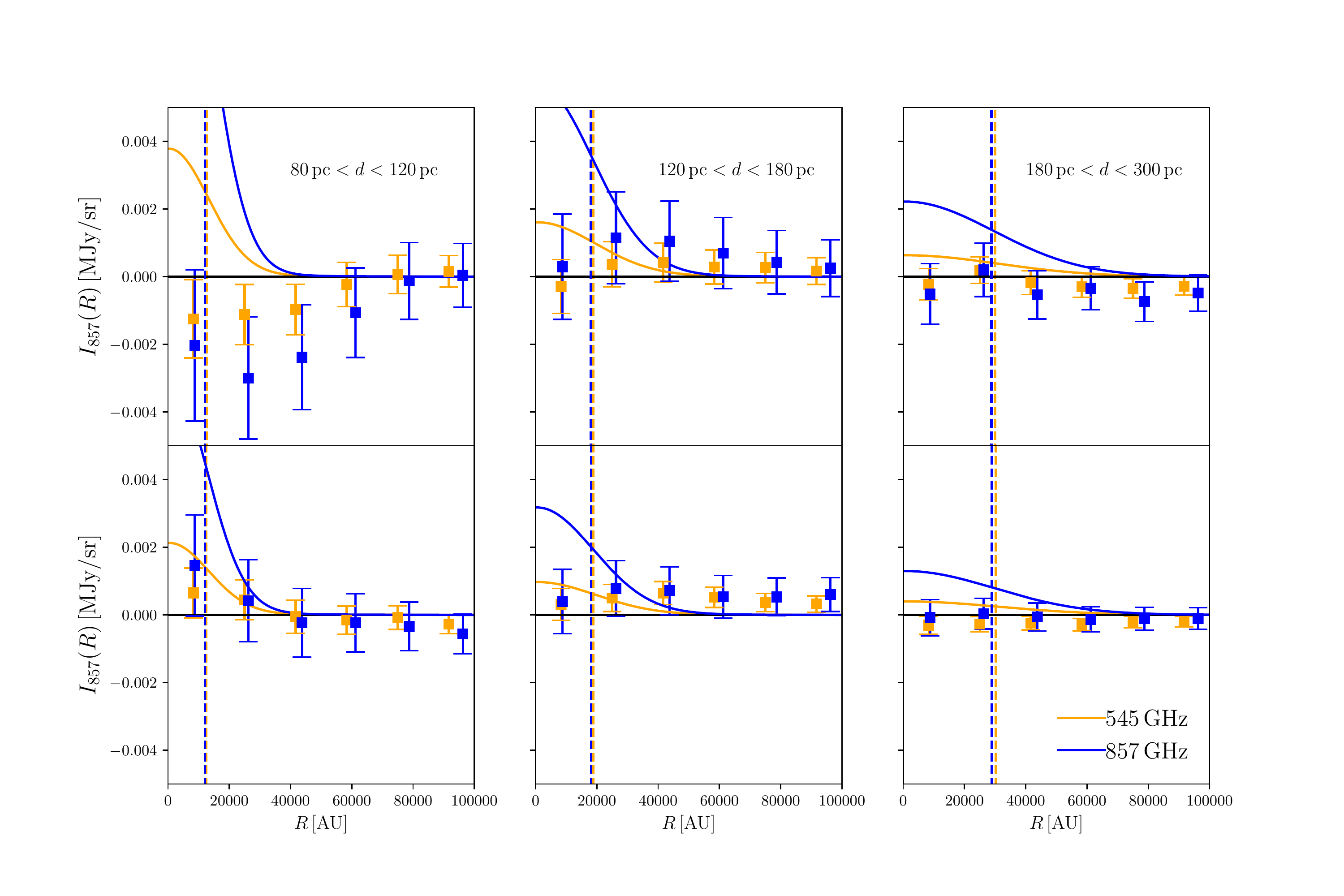}
\caption{\label{fig:measurement} Measurements of 545 GHz (orange) and 847 GHz (blue) signal around {\it Gaia} stars in bins of distance (left to right) and absolute G-band magnitude (top: $2 < M_G < 4$, bottom: $4 < M_G < 6$).  Error bars correspond to the square root of the diagonal of the covariance matrix for each bin; the corresponding correlation matrices are shown in Fig~\ref{fig:correlation_matrix_545} (545 GHz) and Fig~\ref{fig:correlation_matrix_857} (857 GHz).  The number of stars in each bin ranges from $\sim 1400$ for the brightest, closest bin, to $\sim 4\times 10^4$ for the faintest and farthest bin.  Vertical lines indicate the approximate scale of the {\it Planck} beam for each bin and for each channel. The orange (blue) curves show an example EXOC model prediction with $A_M = 50 M_{\oplus}$, $a_{\rm min} = 1.0\,\mu{\rm m}$ and $r_{\rm max} = 10^5\,{\rm AU}$ for the {\it Planck} 545 (857) GHz bands; see text for more details.}
\end{figure*}

\begin{figure}
\centering
\includegraphics[scale=0.55]{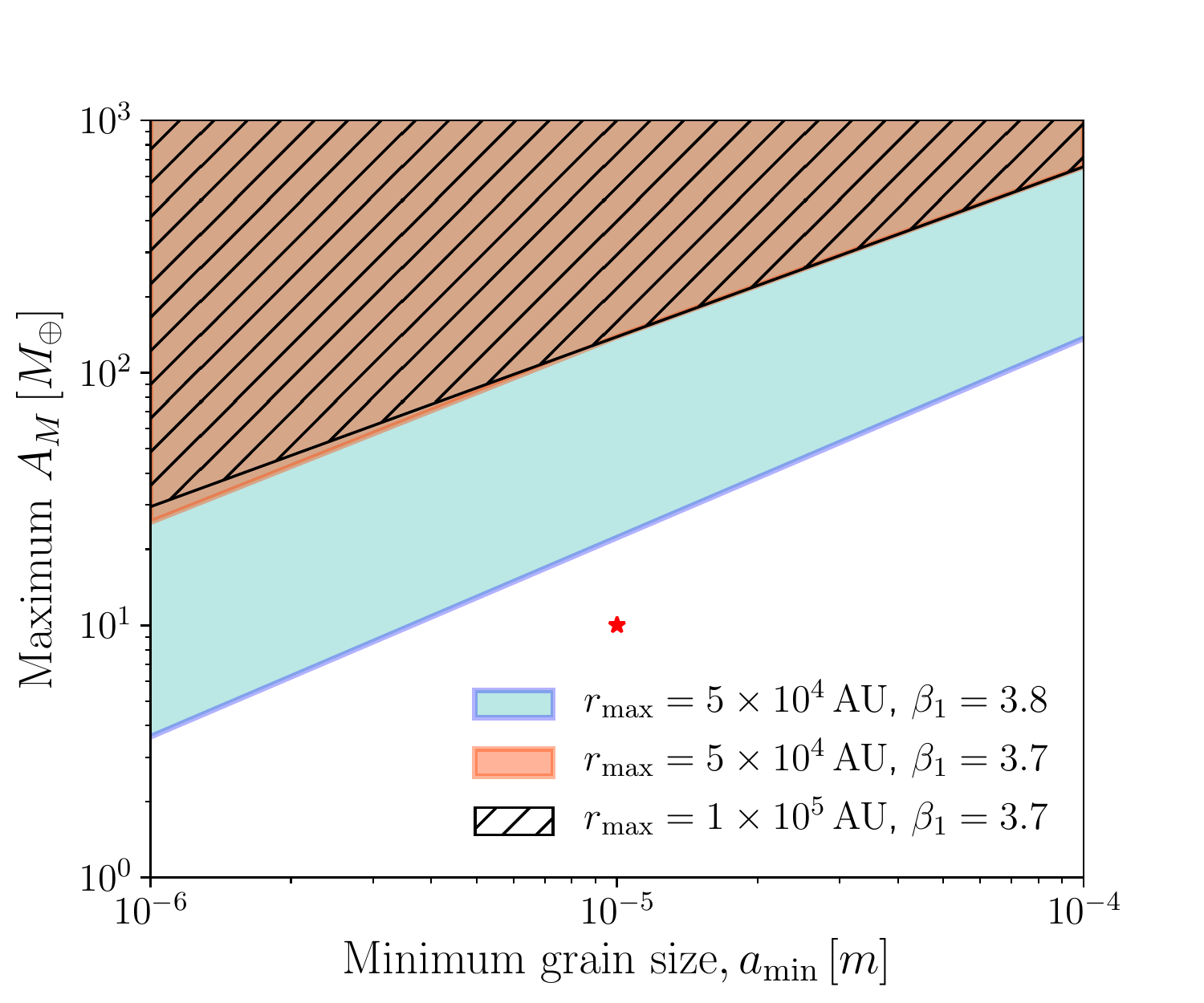}
\caption{\label{fig:exclusion} Limits on the normalization of the Oort cloud mass, $A_M$, as a function of the minimum grain size, $a_{\rm min}$, for different Oort cloud models coming from stacked measurements of the 545 and 857 GHz {\it Planck} maps around {\it Gaia} stars.  Values of $A_M$ in the filled regions are excluded based on the stacked measurements shown in Fig.~\ref{fig:measurement}, assuming that every star in each distance and magnitude bin hosts an identical Oort cloud.  The parameter $r_{\rm max}$ controls the maximum size of the Oort cloud (Eq.~\ref{eq:rho}), while $\beta_1$ controls the size distribution of Oort cloud objects in the small scale regime (Eq.~\ref{eq:size_distribution_small}).  The red star indicates a reasonable estimate for our own Oort cloud.}
\end{figure}

The averaged 545 and 857 GHz signals (orange and blue points with error bars) measured around the stellar sample are shown in Fig.~\ref{fig:measurement} for two different bins of absolute G-band magnitude (top and bottom) and for three bins of stellar distance (left to right).  For illustration, we also indicate the scale corresponding to the {\it Planck} $\sigma_{\rm beam}$ with a vertical dashed line (note that the beam is slightly different for the 545 GHz and 857 GHz maps).  

Fig.~\ref{fig:measurement} shows that the stacked measurements appear consistent with no signal.  We now use these measurements to put constraints on the properties of the EXOCs.  The parameters which most impact the predicted EXOC signal are the normalization of the Oort cloud mass, $A_{M}$ (Eq.~\ref{eq:oort_mass}), the size of the EXOC, $r_{\rm max}$, the power law index of the Oort object size distribution at the small end, $\beta_1$ (Eq.~\ref{eq:size_distribution_small}), and the minimum Oort object size, $a_{\rm min}$.  In Fig.~\ref{fig:measurement} we show the expected EXOC signal for a somewhat extreme EXOC model with $A_{M} = 50\,M_{\oplus}$, $r_{\rm max} = 10^{5}\,{\rm AU}$, and $\beta_1 = 3.7$; the two solid curves show the predicted signal at 545 (orange) and 857 GHz (blue). Note that this model includes the variation in the stellar temperatures and distances in each bin.  As can be seen in the figure, this model is ruled out by the measured signals, assuming that every star hosts an Oort cloud with these properties.  

\subsection{Upper limits on EXOC parameters}

We use the stacked 857 GHz measurements to put limits on the Oort cloud properties in the following manner.  We compute the range of $A_M$ allowed by the data as a function of the assumed minimum grain size, $a_{\rm min}$.  We define a model for the observations:
\begin{eqnarray}
\hat{m}(A_M, a_{\rm min}; d^i, M_g^i, \nu^i, \vec{\theta}) = I(A_M, a_{\rm min}; d^i, M_g^i, \nu^i, \vec{\theta}), \nonumber \\
\end{eqnarray}
where $I$ is the intensity model computed as described in \S\ref{sec:model}, $d^i$ describes the distance bin, $M_g^i$ describes the magnitude bin, $\nu^i$ describes the frequency (either 545 or 857 GHz), and $\vec{\theta}$ represents all other model parameters.  We define a $\chi^2$ for this model relative to the measurements via
\begin{multline}
\chi^2 (A_M, a_{\rm min}) = \sum_{i,j,k} \left(m - \hat{m} (A_M, a_{\rm min}; d^i, M_g^j, \nu^k,\vec{\theta}) \right)^T \\
\mathbf{C}^{-1}  \left(m - \hat{m} (A_M, a_{\rm min}; d^i, M_g^j, \nu^k, \vec{\theta}) \right),
\end{multline}
where $\mathbf{C}$ is the covariance matrix of the observations in a bin.  For each $a_{\rm min}$, we compute the maximum value of $A_M$ such that the minimum $\chi^2$ is less than 49.8, corresponding to a $2\sigma$ upper limit for 35 degrees of freedom.  Repeating this process as a function of $a_{\rm min}$ leads to the upper limits shown in Fig.~\ref{fig:exclusion}.    Note this procedure ignores potential contributions to the measurements from sources of emission other than the EXOCs, including debris disks and the Kuiper belt.  However, including some prescription for these sources of emission in our analysis would only strengthen our limits.   

Fig.~\ref{fig:exclusion} shows the maximum values of $A_M$ allowed (at $2\sigma$ confidence) as a function of $a_{\rm min}$, for several parameter choices.  The red star in the figure illustrates a reasonable estimate for the model parameters of our own Oort cloud.  We do not quite exclude such Oort cloud models.  However, for small $a_{\rm min}$, our limit on the EXOC mass normalization is below the expectation for our own Oort cloud.  For high values of $\beta_1$, our limits on the EXOC mass are significantly stronger owing to more of the mass in the Oort cloud being in smaller objects, which contribute the most to the thermal emission of the cloud.

\subsection{Variations around fiducial analysis}
\label{sec:variations_stacking}

We have made several analysis choices above that could in principle affect the stacked measurements shown in Fig.~\ref{fig:measurement}.  Here we comment on some of the other choices we have explored, and their impact on the results.  

Above, we have restricted the analysis to parts of the sky for which the HI column density is in the lowest 20$\%$ across the sky.  This choice ensures that the stars in our analysis do not live near regions of significant diffuse dust emission that could add significant variance to our measurements, but still allows for the inclusion of a significant number of stars.  We find that making the mask more conservative by using only those regions that have HI column densities in the lowest 10$\%$ across the sky has little impact on the mean of our measurements, but results in an increase to our error bars, as expected by the reduction in the number of stars.  In the other direction, the error bars on the stacked measurements could likely be reduced somewhat by using a less conservative mask, but the gains are not large since the variance in the {\it Planck} maps increases significantly as one moves to regions with higher HI column density.

We have also experimented with changing the sizes of the apertures used when calculating the background flux and with changing the percentile thresholds used when determining the mean flux in the aperture.  We find that prior to subtraction of the signal around random points, making these adjustments results in small constant additive shifts to the measurements that are comparable to the sizes of the error bars.  However, there is no significant impact to the shape of the measured signals.  After performing the random subtraction, this sensitivity to the aperture choices is reduced, suggesting that there may be some coupling of the mask to the measurements of the background.  

The present analysis receives significant noise contributions from galactic backgrounds.  An alternative to using the {\it Planck} maps themselves to estimate these backgrounds, one could imagine using a template for the backgrounds derived in some other way.  We have experimented with using the HI column density map described above in this manner, i.e. using the spatial variation of this map to form an estimate of the spatial variation of the galactic backgrounds to our signal.  However, we find that the resolution of that map is not sufficient to capture some of the small-scale galactic background fluctuations observed in the {\it Planck} maps, and consequently we do not explore this possibility further.

\section{Measurements around individual stars}
\label{sec:hot_results}

The results presented in \S\ref{sec:stacked_results} were found by averaging the 545 and 857 GHz intensity measurements across many stars.  The averaging approach is useful because it can help to beat down noise sources in the measurements.  However, if not every star hosts an Oort cloud, it could miss potential signals and possibly result in artificially tight limits.  In this section, we perform measurements around individual stars. 

\begin{figure}
\centering
\includegraphics[scale=0.55]{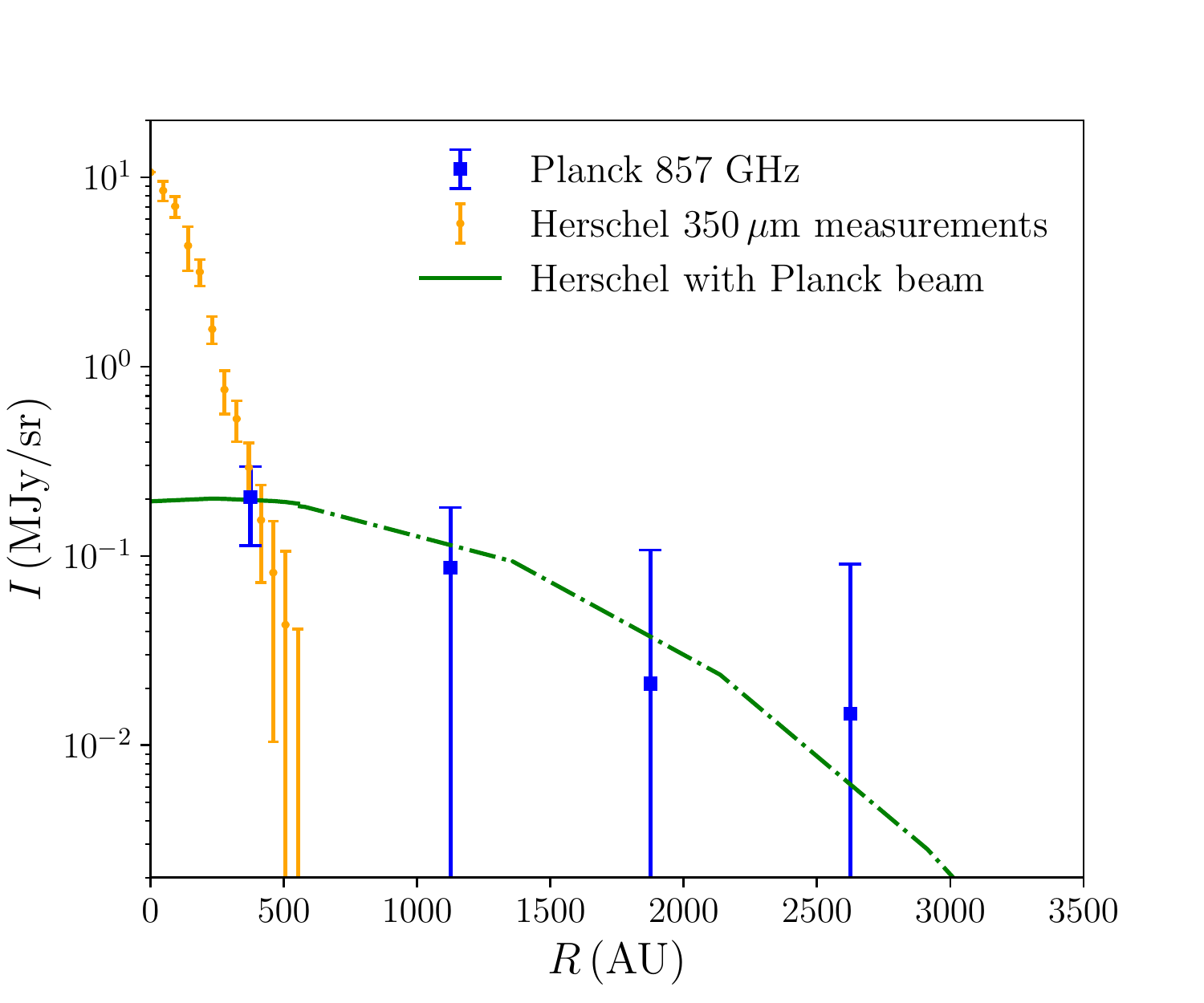}
\caption{Measured {\it Planck} 857 GHz signal around Vega (blue points) compared to measurements from {\it Herschel} (orange points).  Green curve shows the result of applying the {\it Planck} beam to the {\it Herschel} measurements; the dashed region corresponds to an extrapolation assuming that there is no flux from Vega beyond that measured by {\it Herschel}. }
\label{fig:vega}
\end{figure}

\subsection{Vega and Fomalhaut}
\label{sec:vega_fomalhaut}

Two nearby stars known to host debris disks are Vega and Fomalhaut.  Because of these debris disks, the two stars are strong emitters at submillimeter frequencies.  The debris disk of Vega, which lives at 7.68~pc from Earth, has been measured extensively with {\it Herschel} \citep{Sibthorpe:2010}.  Unfortunately, Vega has a low galactic latitude ($19.3^{\circ}$) making it difficult to separate its emission from diffuse backgrounds.  The debris disk of Fomalhaut has also been extensively measured by {\it Herschel} \citep{Acke:2012} and ALMA \citep{MacGregor:2017}, among others.  Fomalhaut is roughly 7.7~pc distant from Earth (by coincidence almost exactly as far away as Vega), but lives at much lower galactic latitude and is therefore in a much quieter part of the sky in the 857 GHz {\it Planck} band.  We measure the 857 GHz {\it Planck} signal around both stars.  These measurements serve two purposes: first, as a check on our measurement approach since both stars have extensive flux measurements in the literature, and second, as targets for detecting extended emission from Oort clouds.  We note that \citet{Stern:1991} also used measurements around individual stars in IRAS data to place limits on their Oort clouds.

A comparison of our measurements around Vega using the 857 GHz {\it Planck} map (blue points) to the {\it Herschel} $350~\mu{\rm m}$ measurements from \citet{Sibthorpe:2010} (orange points) is shown in Fig.~\ref{fig:vega}.  We use the same measurement procedure as described in \S\ref{sec:measurements}, except that we use a smaller range of radial bins since these stars are so close to Earth.  We note that changing the size of the background apertures used for these measurements can change the asymptotic value of the intensity at large $R$, but does not impact the shape of the measured signal.  

Estimating the uncertainty associated with the measurements around Vega is complicated by nearby large background fluctuations.  We take the approach of estimating the uncertainty with a spatial jackknife.  The azimuthally averaged measurements are repeated 20 times, each time with a different patch (we refer to these as jackknife patches) of the map near Vega removed; we use constant angle wedges centered on Vega as the jackknife patches.  The covariance matrix is then estimated from these jackknifed measurements using the standard estimator \citep[e.g.][]{Norberg:2009}.  While this approach captures some of the background fluctuations in the neighborhood of Vega, it misses any large wavelength modes that may be constant over the region considered.  

We compare the measurements around Vega with the 857 GHz band from {\it Planck} to those with the $350\,\mu{\rm m}$ band from {\it Herschel}; these two bands are very well matched in terms of frequency.  Direct comparison of the  {\it Planck} and {\it Herschel} measurements is complicated, however, by the fact that the beam sizes of the two telescopes are very different (roughly 2 arcminutes for {\it Planck} vs. 10 arcseconds for {\it Herschel}).  In order to facilitate comparison between the two measurements, we show the result of convolving the {\it Herschel} measurements with the {\it Planck} 857 GHz beam (green curve).  The solid part of the curve illustrates the region for which we have {\it Herschel} measurements, while the dashed green line is an extension to larger separations assuming that the {\it Herschel} measurements capture all emission from Vega.  We find that the {\it Planck} and {\it Herschel} measurements are completely consistent after taking the differing beam sizes into account.   However, we find that the signal-to-noise of the {\it Planck} measurements is not sufficient to detect or rule out extended emission around Vega.  We do not attempt to constrain possible EXOC emission around Vega because of the complications involved with estimating measurement uncertainty in a region with large galactic background fluctuations.

\begin{figure}
\centering
\begin{tabular}{c}
\includegraphics[scale=0.6]{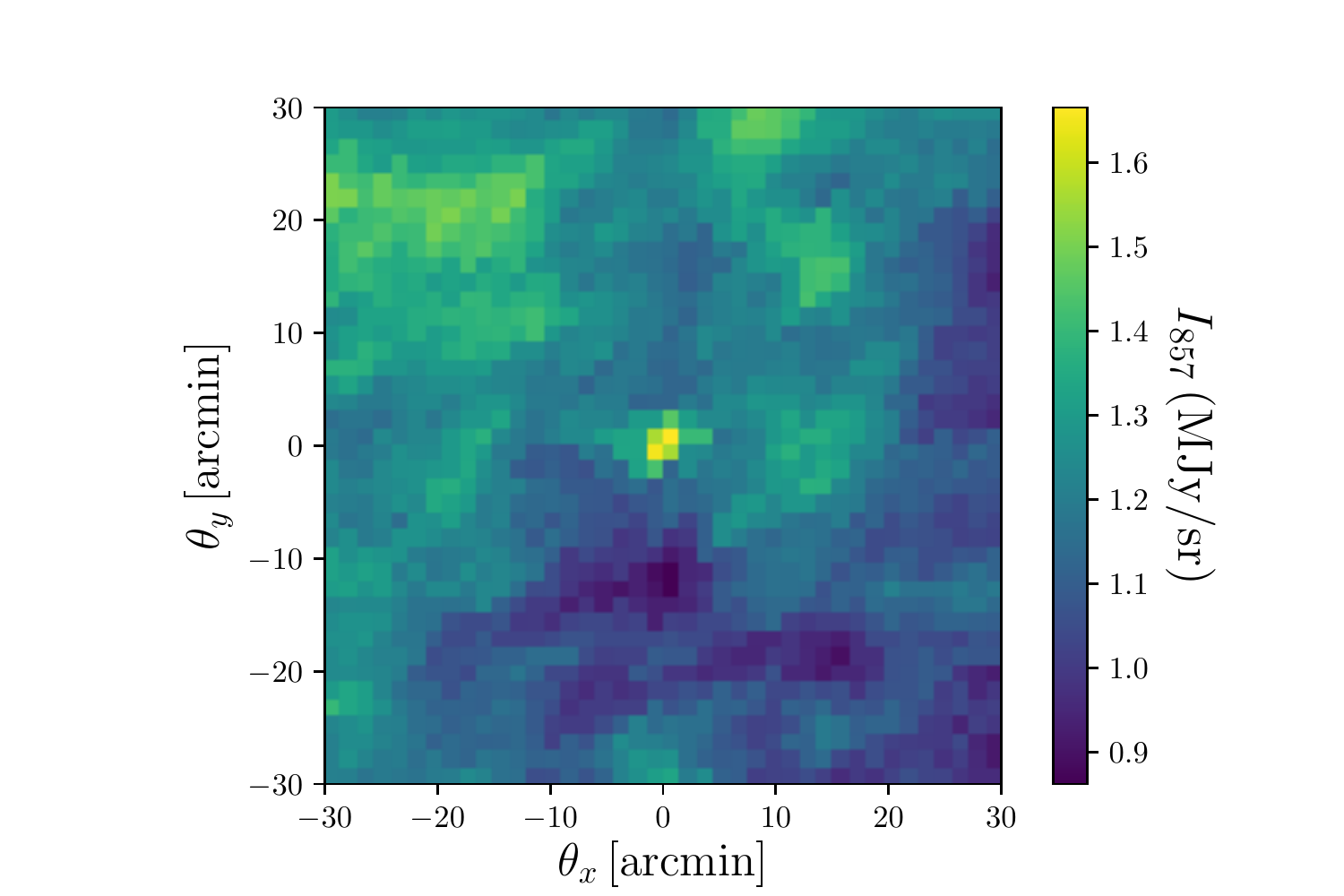} \\
\includegraphics[scale=0.55]{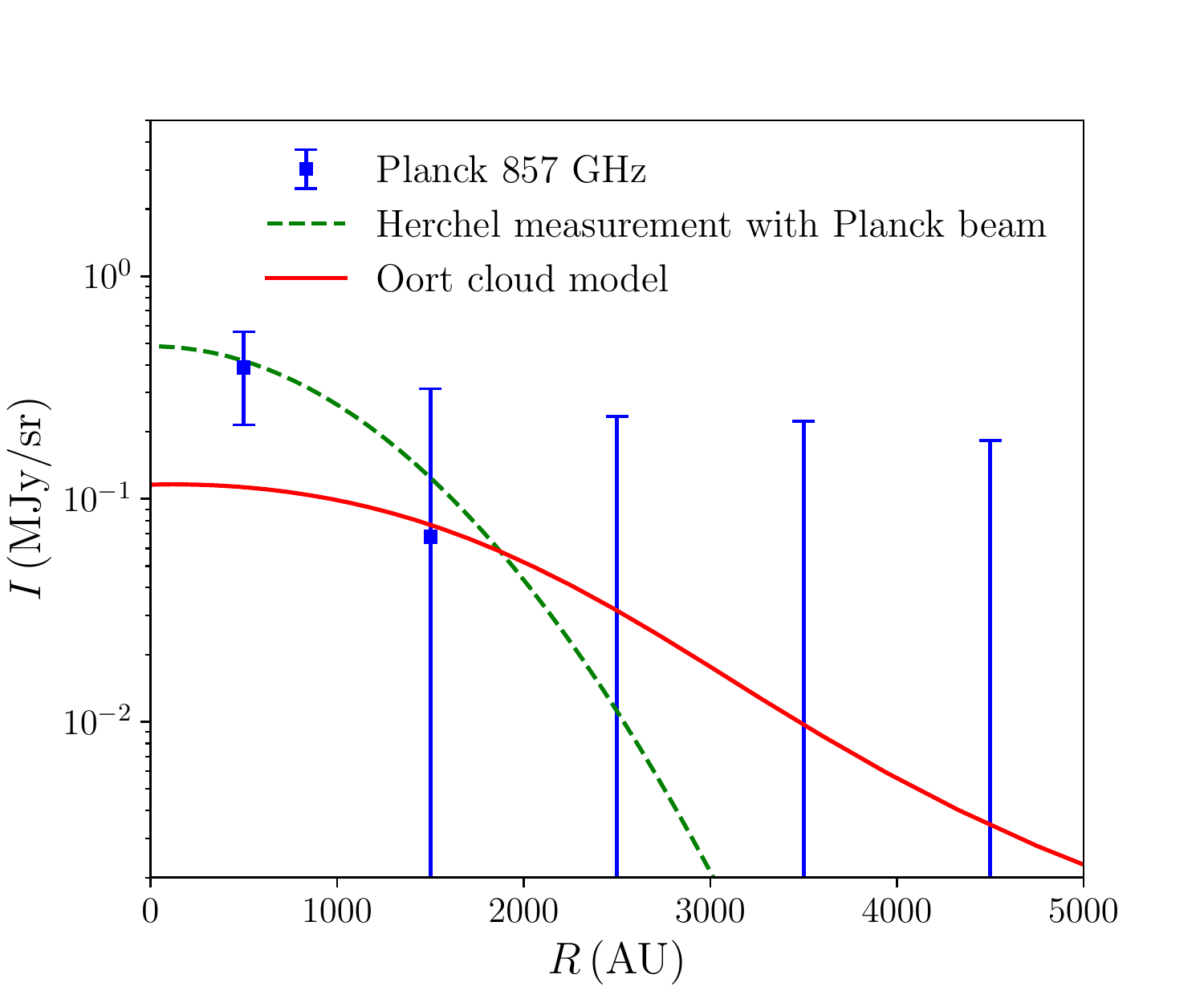} 
\end{tabular}
\caption{Top panel: map of {\it Planck} 857 GHz signal centered on Fomalhaut.  Error bars shown take into account variations due to large scale fluctuations in the galactic backgrounds, as described in \S\ref{sec:vega_fomalhaut}.  Bottom panel: azimuthally averaged 857 GHz measurements as a function of distance from Fomalhaut (blue points).  The green dashed curve illustrates the {\it Planck} beam normalized such that the total intensity matches measurements from {\it Herschel}, while the red solid curve illustrates a fairly typical Oort cloud model, with $A_M = 10 M_{\oplus}$, $a_{\rm min} = 10\,\mu{\rm m}$, $\beta_1 = 3.7$, and $r_{\rm max} = 5\times 10^4\,{\rm AU}$.  Typical exo-Oort cloud emission from Fomalhaut would extend well beyond the beam of {\it Planck}.}
\label{fig:fomalhaut}
\end{figure}

\begin{figure}
\centering
\includegraphics[scale=0.55]{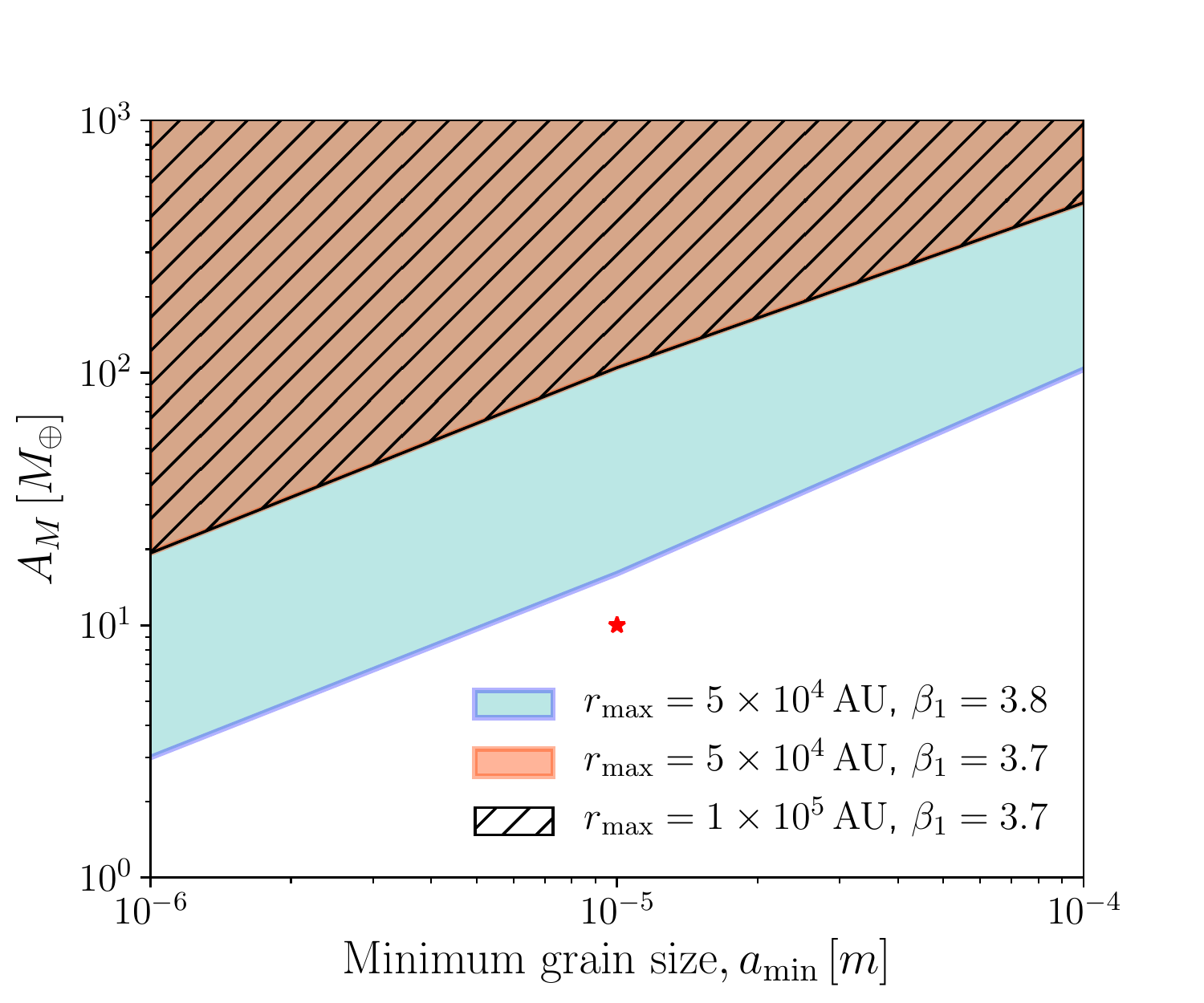}
\caption{\label{fig:exclusion_fomalhaut} Same as Fig.~\ref{fig:exclusion}, except that limits correspond to a potential EXOC around the star Fomalhaut, derived using the measurements shown in Fig.~\ref{fig:fomalhaut}.}
\end{figure}

We repeat the process described above to measure the emission signal around Fomalhaut.  Unlike Vega, Fomalhaut lives in a relatively clean part of the sky.  As a result, its emission can be seen clearly by eye in the {\it Planck} 857 GHz maps (top panel).  The azimuthally averaged measurements around Fomalhaut are shown in the bottom panel of  Fig.~\ref{fig:fomalhaut}.  Because of the fairly uniform galactic background near Fomalhaut, we estimate the uncertainties on our Fomalhaut measurement by repeating these measurements on random points in a 5 deg. $\times$ 5 deg. patch around the star.  The uncertainty is then taken from the covariance across these random point measurements.  This approach has the advantage that it captures  variations in the galactic backgrounds across the map (unlike the jackknife approach used to estimate the uncertainties on our Vega measurements).  However, this approach may result in overestimation of the uncertainties if Fomalhaut happens to live a relatively clean part of the 5 deg. $\times$ 5 deg. patch.  We adopt this somewhat conservative approach, as we will use the Fomalhaut measurements to place upper limits on its EXOC properties below.

We compare our measurements around Fomalhaut using {\it Planck} data to measurements made with the {\it Herschel} 350~$\mu$m band by \citet{Acke:2012}.  As with Vega, we must account for the large beam size of {\it Planck} in this comparison.  The green dashed curve in Fig.~\ref{fig:fomalhaut} shows the {\it Planck} beam, normalized to match the total intensity of Fomalhaut as measured by \citet{Acke:2012}.  We find that our measurements with {\it Planck} are in excellent agreement with those using {\it Herschel} data.   We also plot a typical Oort cloud model (red solid curve) with $A_M = 10 M_{\oplus}$, $a_{\rm min} = 10\,\mu{\rm m}$, $\beta_1 = 3.7$, and $r_{\rm max} = 5\times 10^4\,{\rm AU}$.  The Fomalhaut system is sufficiently close and bright that its expected EXOC emission extends well beyond the {\it Planck} beam.  As with Vega, we do not find evidence for extended emission around Fomalhaut given the large errorbars.  We now proceed to use these measurements to place limits on a potential EXOC around Fomalhaut.

Following the same approach as described in \S\ref{sec:stacked_results}, we use the Fomalhaut measurements to constrain $A_M$ as a function of $a_{\rm min}$.  These results are shown in Fig.~\ref{fig:exclusion_fomalhaut}.  The limits placed on the properties of a Fomalhaut EXOC are comparable to those found in the stacking analysis of \S\ref{sec:stacked_results}.  Note that \citet{Stern:1991} obtained a limit on the optical depth of an EXOC around Fomalhaut of $\tau < 1.5 \times 10^{-5}$ at 100~$\mu$m.

\begin{figure}
\centering
\includegraphics[scale=0.5]{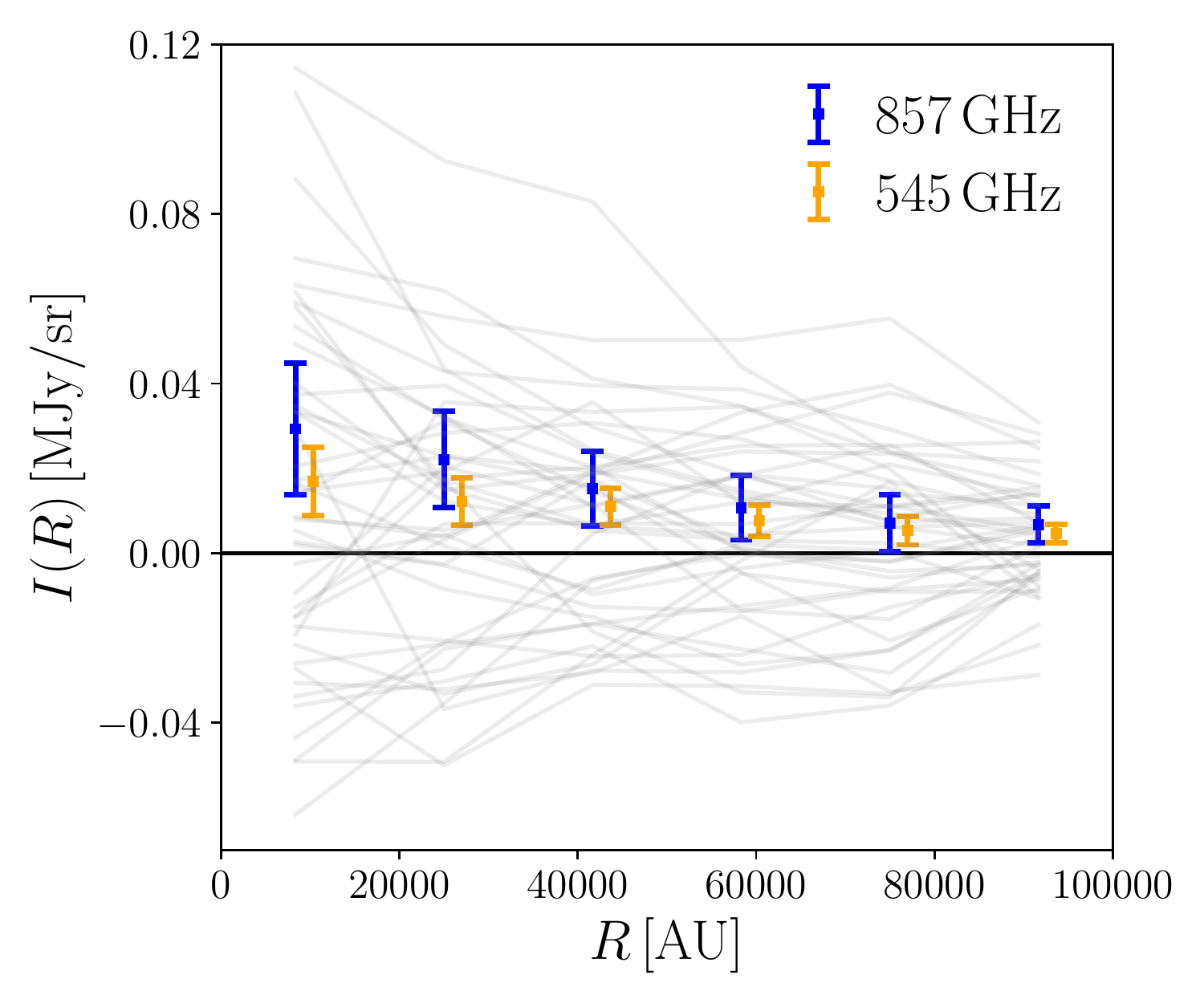}
\caption{\label{fig:bright_stars} Measurements at 545 (orange) and 857 GHz (blue) around 43 `hot' stars with distances between $40 \,{\rm pc}< d < 80\,{\rm pc}$ and effective temperature $T>8000$ K.  Points with error bars show average across all of the hot stars, while gray curves show measurements around individual stars at 545 GHz.  The beam size ($\sigma_{\rm beam}$) corresponds roughly to $10^4\,{\rm AU}$.} 
\end{figure}

\begin{figure*}
\centering
\includegraphics[scale=0.62]{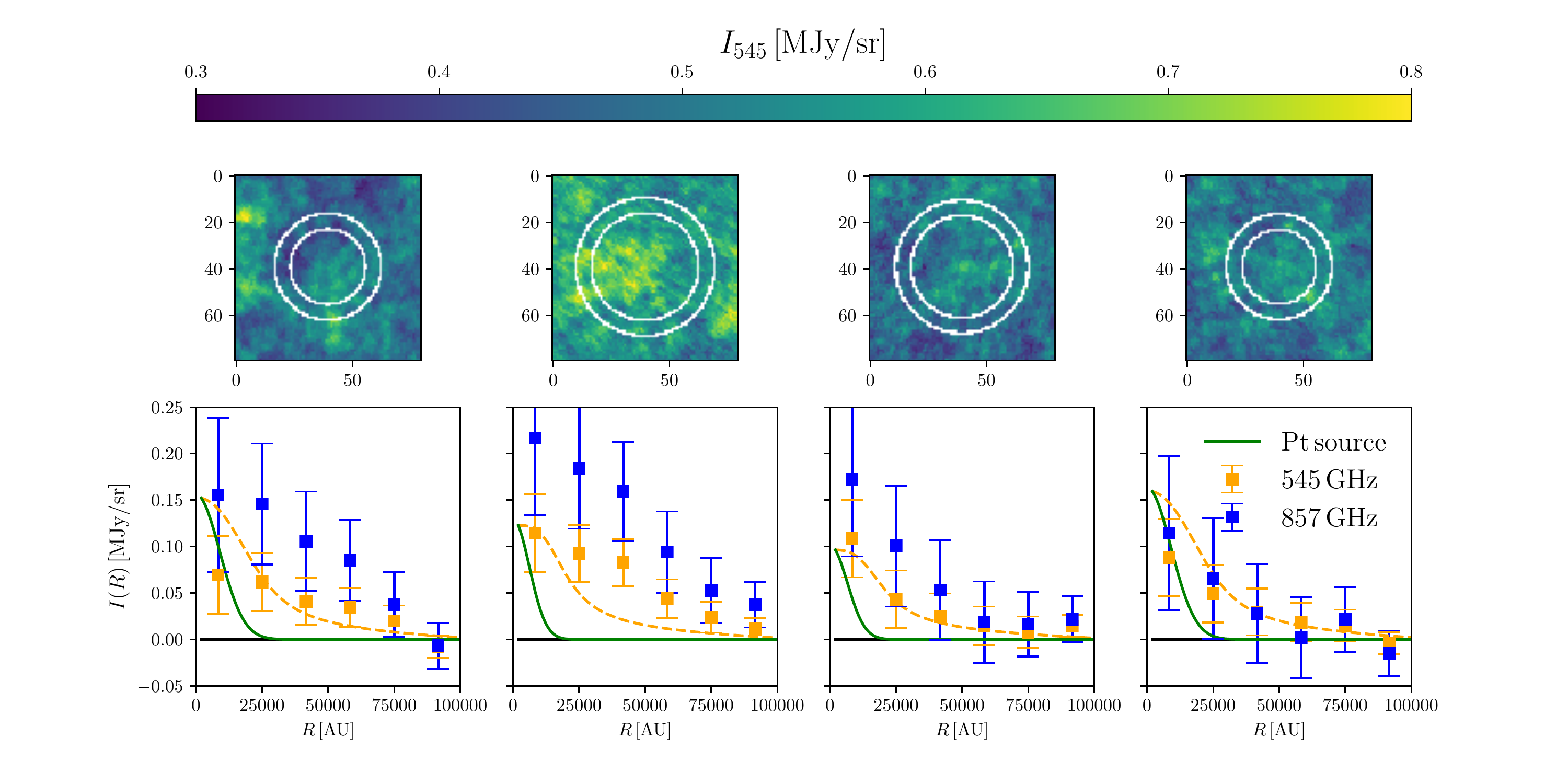}
\caption{\label{fig:indiv_stars} Measured 545 and 857 GHz signal around individual {\it Gaia} stars with $T> 8000\,K$ that show excess signals in the 545 GHz maps.  The upper panels show the 545 GHz maps at the locations of the stars; bottom panels show azimuthally averaged signal for both 545 (orange) and 857 (blue) GHz maps, with error bars derived from the variance across many independent patches within the mask.  We have selected the stars showing the largest excess for this plot.  The dashed orange curve shows a potential EXOC contribution at 545 GHz with a minimum radius, as described in the text.  Parameters of the model have been chosen to roughly match some of the observed excesses.  The green curves in the lower panels show the contribution from a point source for illustration.  The measured signals clearly extend beyond the point source curves, as do the EXOC models. }
\end{figure*}

\begin{figure}
\centering
\includegraphics[scale=0.55]{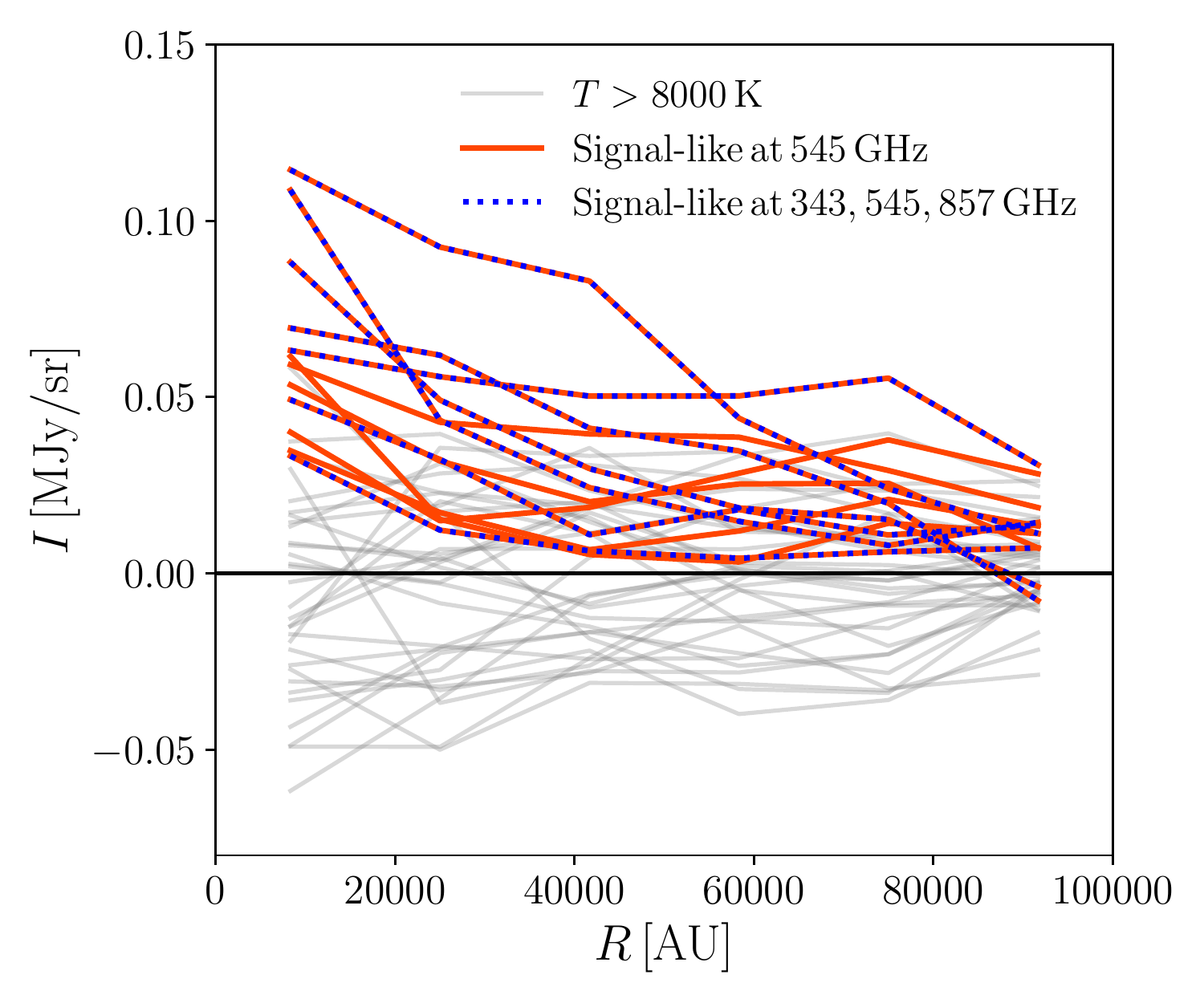}
\caption{\label{fig:hot_star_profiles} Measurements at 545 GHz for individual stars with $T > 8000\,{\rm K}$.  Gray curves show measurements for all such stars, orange curves show measurements for those stars with a signal-like excess (see text for how this is defined), and blue dotted curves indicate those stars for which a signal-like excess is found at 343, 545 and 857 GHz.} 
\end{figure}

\subsection{The EXOCs of nearby hot stars}
\label{sec:hot_stars}

In \S\ref{sec:stacked_results} we restricted the stacked analysis to stars with absolute $M_G > 2$.   We now perform measurements around a sample of nearby bright stars, investigating them both individually and using a stacking approach.

For $M_G \lesssim 2$, the correlation between $M_G$ and the total bolometric luminosity is not very strong.  Consequently, rather than making a selection based on $M_G$, we select instead on {\it Gaia}-estimated temperature, which we expect to be strongly correlated with luminosity. The {\it Gaia} temperature estimates, which are based on three-band photometry, have an internal uncertainty of up to 325~K (\citealt{Andrae:2018}). We select all stars with $T > 8000\,{\rm K}$ and with distances $40 < d < 80\,{\rm pc}$.  The minimum distance cut ensures that large scale structure in the galactic emission does not bias our measurements, but removes only a small number of stars.  We adopt a maximum distance of 80~pc for this analysis so that the expected signal per star is high and the Oort clouds are well resolved by the {\it Planck} beam.  

The measurements around this subset of 43 stars are shown in Fig.~\ref{fig:bright_stars}, with the gray lines indicating the individual measurements in the 545 GHz channel and the points with error bars indicating the averages of the measurements at both 545 and 857 GHz.  The averaged signal shows an excess, with signal-to-noise of $3.2$ for the 545 GHz data and $2.4$ for the 857 GHz data.  Note, however, that there is significant variation from star to star, as illustrated by the gray curves in Fig.~\ref{fig:bright_stars}.

Since there may be physical effects driving considerable variation in EXOC and debris disk emission from star to star, we also explore the measurements around individual stars with $40\,{\rm pc} < d < 80\,{\rm pc}$.  The measurements around the four stars showing the largest excesses at 545 GHz are presented in Fig.~\ref{fig:indiv_stars}.  The top panels show the sky signal in $120' \times 120'$ patches around each of these stars, while the bottom panels show the azimuthally averaged signal.  The white circles in the top panels indicate the annuli used to measure the background for each star; they vary in size because the stars are at different distances.  To determine the error bars for the azimuthal measurements around individual stars, we perform the same measurements around random points and compute the variance across these measurements.  The error bars therefore reflect variation in the maps due to fluctuating backgrounds.

The measurements around the largest excesses appear to have signals with shapes roughly consistent with expectations for EXOC-like emission.  Assessing the statistical significance of these measurements is complicated by at least three factors: (a) we have selected stars that show large excesses, (b) we do not necessarily expect a signal around every star, and (c) the distribution of the background flux is non-Gaussian.  Indeed, measurements around random points occasionally show signals similar to those in Fig.~\ref{fig:indiv_stars}.  

To assess the significance of these measurements, we first define what we would consider a potentially interesting signal.  As illustrated in Fig.~\ref{fig:model_curves}, we expect the signal from an EXOC to be a steeply declining function of radial distance.  We therefore define interesting signals with the following two criteria:
\begin{itemize}
\item The signal is positive in the smallest three radial bins.  This ensures that the emission is extended.
\item The signal in the first radial bin is larger than that in all other radial bins.  This ensures that the signal is decreasing as a function of distance from the star, which is a robust prediction of our models.
\end{itemize}
These criteria are quite general, and should encapsulate most reasonable Oort cloud signals.  Of course, noise fluctuations may cause a true Oort cloud signal to not meet these selection criteria.  We will explore variations in these criteria below.

We  find that 12\% of measurements around random points (after applying the same masking cuts as for the real stars) meet these criteria.  We therefore conclude that given these selection criteria, our false positive rate for this selection is 12\%.  Using the actual stars locations, though, we find that 12 out of the 43 stars in the bright sample, or $28\%$, meet the selection criteria.   

The profiles of the $T > 8000\,{\rm K}$ stars identified as having signal-like excesses at 545 GHz are shown as the orange curves in Fig.~\ref{fig:hot_star_profiles}.  Also shown (grey curves) are the profiles of stars that do not meet our criteria for having signal-like excesses.  As in Fig.~\ref{fig:indiv_stars}, the {\it Planck} beam corresponds roughly to physical scales of 10000 AU in this plot.  The orange curves in Fig.~\ref{fig:hot_star_profiles} appear to be significantly more extended than point source emission.

We can determine whether 12 out of 43 stars meeting the selection criteria is significant in the following way.  Since there is a 12\% chance of meeting the selection criteria by chance, the probability that 12 out of 43 stars happen to meet the selection criteria by chance is given by the binomial distribution:
\begin{eqnarray}
P(12) &=& \mathcal{B}(k=12,n=43,p=0.12) \nonumber \\
&=& \binom{n}{k} p^k(1-p)^{n-k}.
\end{eqnarray}
The chances of 12 or more stars meeting the selection criteria by chance is then
\begin{eqnarray}
p &=& 1-\sum_{i=0}^{i=11} \mathcal{B}(k = i,n = 43,p = 0.12) \\
&=& 0.005.
\end{eqnarray}
We therefore conclude that the fact that 12 stars out of 43 meet our criteria for potential signals is statistically significant.  However, we caution that the false positive rate is very high.  Of the measurement in the sample, we expect $pn=0.12\times 43 = 5.2$ of them to occur by random chance alone.  

The excesses shown in  Fig.~\ref{fig:indiv_stars} are significantly more extended than point source emission.  We illustrate a point source signal convolved with the {\it Planck} beam with the blue curve.  Also shown in Fig.~\ref{fig:indiv_stars} is an example of an EXOC cloud model that can roughly fit some of the measurements.  As seen in the figure, the Oort cloud model yields emission that is more extended than a point source (the point source amplitude is scaled to match that of the EXOC model at $R=0$).  Fitting the measurements requires a somewhat extreme choice of model parameters.  The model shown in Fig.~\ref{fig:indiv_stars} has $A_M = 10^3 M_{\oplus}$, $a_{\rm min} = 1.0\,\mu {\rm m}$, $\beta_1 = 3.8$, $\gamma = 2.0$, $r_{\rm min} = 10^{4}\,{\rm AU}$ and $r_{\rm max} = 10^{5}\,{\rm AU}$.  This model is both more extended and more luminous than would be naively expected based on constraints from our own Oort cloud.  We emphasize, though, that the properties of EXOCs are largely unknown, and it is possible that some fraction of  bright stars have larger Oort clouds than would be expected from a simple extrapolation of the properties of our own Oort cloud.  Such EXOC models would also be ruled out by the constraints from our own stacking analysis shown in Fig.~\ref{fig:exclusion}.  However, as discussed previously, those limits assume that every star hosts an identical EXOC.  If most stars do not host an EXOC, then the limits shown in Fig.~\ref{fig:exclusion} would not rule out the possibility of the hottest stars hosting extreme EXOCs.  We discuss below other possible sources of thermal emission at large distances from host stars. 

\subsection{Properties of the hot star sample}

As shown by the locations of the 43 stars in our hot, nearby star sample in the HR diagram in Figure \ref{fig:star_selection}, we find that most of these objects are A dwarf stars (35 of 43), with the rest being slightly evolved A stars or main sequence B stars. As expected based on the overall statistics of the A star population, a significant fraction of these stars are known binaries (24 of 43). Similarly, again consistent with the overall properties of the A star population, several of the stars in our sample are reported to have chemical peculiarities in their spectra (8 of 43). We find no significant differences between the average properties of the stars in our sample with and without excess 857 GHz and 545 GHz emission. We considered the ages of these stars based on the work of \citet{David:2015}, who used narrow-band photometry compared to isochrones to estimate ages of early type stars. We find that 38 of the stars in the sample have age estimates and that those ages span a range of 65 to 725~Myr. The average age of the of the 12 objects with potential excess emission is 247$\pm$52~Myr, while the average age of the remaining stars is 350$\pm$40~Myr, indicating that the potential excess stars are possibly younger. We note that we find no very young objects ($\approx$10~Myr) where emission phenomena specific to Young Stellar Objects may be expected.  

A significant fraction of A stars are known to have debris disks, and the rate of occurrence of these disks is observed to decline steeply at ages greater than 300 to 400 ~Myr (\citealt{Habing:1999}). Some of the stars in our hot, nearby star sample are known to have debris disks. We cross matched the stars in our sample with known, resolved debris disks\footnote{\url{https://www.circumstellardisks.org}} and found that two stars in our sample have well-studied debris disks - 49 Cet (\citealt{Roberge:2013}), for which we find a 857 GHz excess, and  HD 21997 (\citealt{Kospal:2013}), for which we do not. We also compare our sample to known IR excess stars from \citet{McDonald:2017} and find only one overlap. However, we note that catalogs of IR excess stars are based in part on IRAS measurements at 60 and 100 $\mu$m, which probe significantly hotter thermal emission than the \textit{Planck} data. In addition, many IR excess stars will be evolved or very young stars, which have intentionally been excluded from our analyses. 

In some cases, A stars are known to have substantial magnetic fields, particularly the chemically peculiar Ap stars (see \citealt{Mathys:2017} and references therein). It is possible that electrons in the stellar winds of these stars could produce radio emission that has a large physical extent through gyrosynchrotron radiation (\citealt{Gudel:2002}). Comparison of 857 GHz to 545 and 343 GHz signals in our data argues strongly against this source of radiation since for synchrotron radiation the emission at lower frequencies should be stronger. Similarly, we find that there are no NVSS (\citealt{Condon:1998}) 1.4 GHz radio sources within 30'' of any member of our hot, nearby star sample. While we cannot definitively exclude the possibility of an emission source that is physically associated with the star, but not gravitationally bound to it, this explanation appears unlikely given the available data. If a significant population of high energy electrons are involved in the emission process, we might expect a corresponding X-ray signature due to inverse Compton scattering in the radiation field of the star. We searched the ROSAT point source catalog of \citet{Boller:2016} and found that 11 of 43 stars in our hot, nearby star sample have X-ray point sources within 30''. Of these, four are found to have 857 GHz excesses in our analysis. Given the number of sources in the ROSAT catalog ($\approx1.35\times10^{6}$), the probability of X-ray sources falling within 30'' of any one of the stars in our sample is less than $1\%$.  We note that the ROSAT sources are very non-uniformly distributed in galactic longitude, so this is an overestimate of the probability of ROSAT sources falling within 30'' of our stars. Previous studies of X-rays from A stars have reported a similar fraction of objects with detectable emission (\citealt{Simon:1995}), which they attribute to chromospheric emission in the case of isolated A stars.

Finally, we note that two of the stars in the hot, nearby star sample (one with 857 GHz excess, one without) are known $\alpha2$CVn variable stars. These stars are thought to have strong magnetic fields, rapid rotation, and surface features that evolve in time, causing low-level  photometric variability.

\subsection{Variations around fiducial analysis choices}

One could imagine changing the criteria introduced above for selecting possible excess signals.  Changing the criteria will necessarily change the false positive rate of the selection.  We find, though, that the $p$-values continue to be small for reasonable variations around our fiducial choices.  For instance, if we require the signal to be positive in the first two bins rather than the first three, the $p$-value changes to 0.007; increasing the requirement to the first four bins reduces the $p$-value to $8\times 10^{-4}$.  This change in $p$-values suggests that the random fluctuations have a typically narrower profile than the on-star fluctuations.  Alternatively, if we require that the signal in the first bin be larger than the third through sixth bin (rather than all other bins), then the $p$-value changes to $0.003$.  Finally, using the same signal requirement, but switching to the 857 GHz data results in a $p$-value of $0.007$.  In all cases, the on-star measurements show significantly more EXOC-like signals than the random point measurements.   In Fig.~\ref{fig:hot_star_profiles} we identify those stars (orange curves with blue dots) that show signal-like excess in 343, 545 and 857 GHz. 

We also tested an alternative method where we select stars with lower background contamination, motivated by the analysis of \citet{Stern:1991}.  Apertures around each star were split into halves along lines of constant longitude and latitude.  If the mean signal in two halves was determined to be significantly different, that star would be excluded from the analysis.  The intention of this selection was to remove stars that live on large-scale galactic cirrus fluctuations.  However, we found that this approach could potentially introduce significant selection biases into the measurements, and so we do not present those results here.

Finally, we have also tried changing the temperature selection in this analysis.  We find that as our temperature threshold is reduced from $8000\,{\rm K}$ to below $\sim 7000\,{\rm K}$, the statistical significance of the stacked signal shown in Fig.~\ref{fig:bright_stars} declines.  Note that the number of stars increases rapidly with lower temperature threshold, and these low temperature stars will therefore dominate the average.  It appears that the apparent excess is being driven by the stars with the highest temperatures.

\bigskip 
\bigskip

\section{Discussion}
\label{sec:discussion}

\subsection{Summary and caveats}

The analysis presented here demonstrates that detection of extra solar Oort clouds (EXOCs) with data from CMB surveys is promising.  By correlating the 545 and 857 GHz {\it Planck} maps with {\it Gaia}-detected stars, we place limits on the properties of EXOCs, in particular on the mass contained in them and the minimum grain size (Figs.~\ref{fig:exclusion} and \ref{fig:exclusion_fomalhaut}).  We compare our measurements with known debris disk systems -- in the case of the stars Vega and Fomalhaut we find a significant excess that is in agreement with measurements from {\it Herschel}. With conservative estimates of the uncertainty due to background fluctuations around these stars, we do not see a significant excess at large distances beyond the debris disk signal convolved with the {\it Planck} beam.  We use the measurements around Fomalhaut to constrain a possible EXOC of that system. 

We have also identified a potentially interesting excess emission signal around nearby hot stars, shown in Fig.~\ref{fig:bright_stars} and \ref{fig:indiv_stars}. We found statistically significant signal in the {\it Planck} 857 GHz and 545 GHz channels at distances of  $10^4$ to $10^5$ AU in a sample of 43 nearby hot stars, mostly A stars.  This emission can be fit reasonably well with EXOC models, although doing so requires somewhat extreme parameter choices relative to the constraints on the properties of our own Oort cloud.  Such extreme EXOC models would also be ruled out by the constraints shown in Fig.~\ref{fig:exclusion}, although those limits could be avoided if only a small fraction of stars host EXOCs.

EXOC emission is not the only explanation for the apparent excess emission around our hot star sample.  We have described our procedure for masking and subtracting potential contributions from galactic dust, which is our primary source of systematic uncertainty.  If the signal is indeed real, it could result from at least two other sources: a ``halo'' of particles ejected from the debris disks by radiation pressure or stellar winds, or nebular emission seen in young stars. The ages of the hot stars in our sample are typically larger than 100 Myr, so the latter explanation appears unlikely.  Distinguishing a genuine exo-Oort cloud from a particle ``halo'' appears challenging, as the emission in both cases is dominated by small grains.  We have not pursued these possibilities in any detail but future studies with higher resolution and sensitivity data would be valuable.  

The connection of thermal emission well beyond the scale of debris disks to the planetary system of host stars is a topic of great interest. The generation of extended scattered disks and Oort clouds like our own require the dynamical presence of Neptune-Jupiter sized planets for Sun-like stars. The scale of the Oort cloud is also influenced by the early environment of the star, such as the possible presence of a star cluster. Hence empirical knowledge of the statistics of EXOCs and their correlation with the planets and debris disks of host stars can yield insights on the early stages of the formation of stars and planetary systems. 

\subsection{Strategies for future measurements}

We now turn our attention to prospect for future analyses.  In the current analysis, we have focused on two measurement strategies: stacked measurements on many stars, and measurements around individual bright stars.  The data requirements and aims of future analyses will depend on which measurement strategy is adopted.  The advantage of a stacked analysis is that one can effectively beat down noise sources by averaging.  However, it is not clear whether all stars host Oort clouds, so this averaging may also reduce the measured signal.  Observations of individual stars do not suffer from this drawback, but must reach significantly higher sensitivities per star to reach the threshold for detection.

Both approaches seem worth pursuing in future analyses.  For a stacking analysis, wide field survey data is ideal since it will enable averaging over many stars.   For measurements around individual stars, targeted observations would likely offer higher signal to noise.  

In both cases, one must determine a distance threshold to use for identifying stellar candidates.  For distances $d \lesssim 300\,{\rm pc}$, the number of stars available for the analysis goes as the distance cubed.  At distances of $d \gtrsim 300\,{\rm pc}$, however, one reaches the limits of the disk of the Milky Way, and the number of stars will increase more slowly with distance. Additionally, separating EXOC emission from possible debris disk emission could be very difficult for distant stars.  If the EXOC is resolved (which may be realistic up to around 1~kpc), its surface brightness will be independent of distance.  Even so, the number of beams one can average, which is desirable, is greater for nearby EXOCs.   

Given expected temperature estimates in the range of 10 to 50 K for parts of EXOCs, wavelengths from about 60 to 300 $\mu$m or frequencies from about 600 to 5000 GHz would be well matched to detection of EXOC emission.  For sun-like stars, the lower wavelength and upper frequencies ends of these ranges would be optimal.  While pushing to lower frequencies may not be optimal in terms of the amplitude of the signal, the background from galactic cirrus will also be reduced.  As seen in several of the figures above, the error bars for the 545 GHz measurements are typically smaller than those of the 857 GHz measurements for precisely this reason.

While the expected temperature ranges of EXOCs are well constrained, the signal amplitude is largely uncertain.  For reasonable choices of model parameters, the intensity of the EXOC thermal emission ranges from about $10^{-2}\,{\rm MJy}/{\rm sr}$ in the inner parts of the clouds to $10^{-6}\,{\rm MJy}/{\rm sr}$ in the outer parts  (see e.g. Fig.~\ref{fig:model_curves}).  Sensitivities better than $10^{-5}\,{\rm MJy}/{\rm sr}$ would be ideal for detecting extended EXOC emission, but there are regions of parameter space where detection could be achieved with considerably less sensitive measurements.  

In addition to high sensitivity, perhaps the most desirable feature for future observations would be high angular resolution.  For reasonable model parameters, the EXOC signal drops rapidly with distance from the star.  Consequently, for a large beam, it is difficult to distinguish EXOC emission from point-like emission, which could for example be sourced by a debris disk.  Furthermore, the background fluctuations from galactic dust have less power on very small scales, so a small-scale EXOC signal could be more easily separated from backgrounds.  Pushing the beam to below $R = 10^3\,{\rm AU}$, either by using nearby stars or a higher angular resolution instrument, would be ideal.  

\subsection{Prospects for EXOC measurements using CMB and far-infrared datasets}

In this analysis, we have used 545 and 857 GHz data from \emph{Planck} to attempt to measure EXOC emission.  This dataset is well matched in terms of frequency and has the advantage of large sky coverage.  However, the measurements with \emph{Planck} suffer from the effects of a  large beam relative to the scales of an EXOC.  There are several current and future datasets that may be well matched to detecting Oort cloud emission, which we summarize below.

Given their large sky coverage and high sensitivity to microwave frequencies, CMB survey data is well matched to stacked searches for Oort cloud emission. Among ongoing CMB experiments, the South Pole Telescope (SPT; \citealt{Carlstrom:2011}) and the Atacama Cosmology Telescope (ACT; \citealt{Swetz2011}) offer higher resolution and deeper maps over thousands of square degrees of the sky.

Future CMB observations from Advanced ACTPol \citep{Henderson:2016}, SPT-3G \citep{Benson:2014}, the Simons Observatory \citep{SOforecast} and CMB-S4 \citep{Abazajian:2016} will provide significantly deeper and higher resolution maps of the microwave sky than the {\it Planck} maps considered here, and over wide regions of the sky. 

With current {\it Planck} data we have reached sensitivities of roughly  $10^{-3}\,{\rm MJy}/{\rm sr}$ in stacked measurements.  Future data from stage-III and stage-IV CMB experiments is expected to be roughly 10 and 100 times more sensitive than current \emph{Planck} data, respectively \citep{Abazajian:2016}.   However, a significant limitation of the current analysis is our ability to model galactic backgrounds.  With higher resolution data, background modeling could be significantly improved, and presumably more stars could be included in the analysis.   Assuming a factor of 10 increase in the number of stars and factors of 10 or 100 improvements in the sensitivity, it should be possible to reach sensitivities of roughly $3\times 10^{-6}$ to $10^{-5}\,{\rm MJy}/{\rm sr}$ in stacked measurements.  At these sensitivities, reasonable Oort cloud models could be detected out to their edges, at $R \sim 5\times 10^{4}$ AU.  Additionally, this level of sensitivity should be sufficient to detect emission from individual EXOCs around nearby stars for reasonable models.

Wide-field CMB or infrared surveys are well matched for stacked searches for EXOC emission.  Targeted observations at similar frequencies, on the other hand, could be used to detect emission from individual EXOCs.  There are several current instruments that could potentially be used to this end, including the Large Millimeter Telescope, MUSTANG-2 \citep{Mustang2}, ALMA \citep{ALMA}, and BLASTPol \citep{Blastpol}. 

While the peak emission of a 10K Oort cloud is well matched to the 545 and 857 GHz maps considered here, the inner, hotter parts of EXOCs could have peak emission closer to the infrared bands.  As noted previously, searches for EXOCs have already been attempted using IRAS \citep{Stern:1991}.  Other infrared instruments with capabilities similar to those of {\it Herschel} could  be potentially well matched to EXOC detection. 
Debris disks around stars have already been detected by {\it Herschel}; searching for Oort cloud emission would simply involve extending these searches to larger distances from the parent stars.  Further out in time, the WFIRST survey may provide additional constraints on EXOCs in the infrared, including possible detection of bodies in the Inner Oort cloud \citep{Holler:2017}.

Finally, we expect our understanding of the properties of our own Oort cloud to improve significantly with future observations from the Large Synoptic Sky Survey (LSST) \citep{LSSTsciencebook,Trilling:2018}.  By observing significant numbers of long period comets and Halley-type comets, LSST should constrain dynamical models of the Oort cloud, resulting in improved constraints on e.g. its density profile.  Such constraints will inform searches for EXOCs as well.

\bigskip
\vspace{1cm}

{\it Acknowledgements}

We are grateful to Gary Bernstein, Mark Devlin, Doug Finkbeiner, Mike Jarvis, Renu Malhotra, Tom Crawford, Gil Holder and Adam Lidz for many helpful discussions and feedback on an early draft. We thank James Aguirre, Ana Bovana, Chihway Chang, Neal Dalal, Meredith Hughes, Jeff Klein, Anthony Lewis, Niall MacCrann, Gemma Moran, Eduardo Rozo, Masao Sako, Carles S\'anchez, Masahiro Takada, Martin White for stimulating discussions. 

\bibliography{thebibliography.bib}

\begin{thebibliography}{}
\expandafter\ifx\csname natexlab\endcsname\relax\def\natexlab#1{#1}\fi
\providecommand{\url}[1]{\href{#1}{#1}}

\bibitem[{{Abazajian} {et~al.}(2016){Abazajian}, {Adshead}, {Ahmed}, {Allen},
  {Alonso}, {Arnold}, {Baccigalupi}, {Bartlett}, {Battaglia}, {Benson},
  {Bischoff}, {Borrill}, {Buza}, {Calabrese}, {Caldwell}, {Carlstrom}, {Chang},
  {Crawford}, {Cyr-Racine}, {De Bernardis}, {de Haan}, {di Serego Alighieri},
  {Dunkley}, {Dvorkin}, {Errard}, {Fabbian}, {Feeney}, {Ferraro}, {Filippini},
  {Flauger}, {Fuller}, {Gluscevic}, {Green}, {Grin}, {Grohs}, {Henning},
  {Hill}, {Hlozek}, {Holder}, {Holzapfel}, {Hu}, {Huffenberger}, {Keskitalo},
  {Knox}, {Kosowsky}, {Kovac}, {Kovetz}, {Kuo}, {Kusaka}, {Le Jeune}, {Lee},
  {Lilley}, {Loverde}, {Madhavacheril}, {Mantz}, {Marsh}, {McMahon},
  {Meerburg}, {Meyers}, {Miller}, {Munoz}, {Nguyen}, {Niemack}, {Peloso},
  {Peloton}, {Pogosian}, {Pryke}, {Raveri}, {Reichardt}, {Rocha}, {Rotti},
  {Schaan}, {Schmittfull}, {Scott}, {Sehgal}, {Shandera}, {Sherwin}, {Smith},
  {Sorbo}, {Starkman}, {Story}, {van Engelen}, {Vieira}, {Watson}, {Whitehorn},
  \& {Kimmy Wu}}]{Abazajian:2016}
{Abazajian}, K.~N., {Adshead}, P., {Ahmed}, Z., {et~al.} 2016, ArXiv e-prints,
  arXiv:1610.02743

\bibitem[{{Acke} {et~al.}(2012){Acke}, {Min}, {Dominik}, {Vandenbussche},
  {Sibthorpe}, {Waelkens}, {Olofsson}, {Degroote}, {Smolders}, {Pantin},
  {Barlow}, {Blommaert}, {Brandeker}, {De Meester}, {Dent}, {Exter}, {Di
  Francesco}, {Fridlund}, {Gear}, {Glauser}, {Greaves}, {Harvey}, {Henning},
  {Hogerheijde}, {Holland}, {Huygen}, {Ivison}, {Jean}, {Liseau}, {Naylor},
  {Pilbratt}, {Polehampton}, {Regibo}, {Royer}, {Sicilia-Aguilar}, \&
  {Swinyard}}]{Acke:2012}
{Acke}, B., {Min}, M., {Dominik}, C., {et~al.} 2012, \aap, 540, A125

\bibitem[{{Andrae} {et~al.}(2018){Andrae}, {Fouesneau}, {Creevey}, {Ordenovic},
  {Mary}, {Burlacu}, {Chaoul}, {Jean-Antoine-Piccolo}, {Kordopatis}, {Korn},
  {Lebreton}, {Panem}, {Pichon}, {Thevenin}, {Walmsley}, \&
  {Bailer-Jones}}]{Andrae:2018}
{Andrae}, R., {Fouesneau}, M., {Creevey}, O., {et~al.} 2018, ArXiv e-prints,
  arXiv:1804.09374

\bibitem[{{Anglada} {et~al.}(2017){Anglada}, {Amado}, {Ortiz}, {G{\'o}mez},
  {Mac{\'{\i}}as}, {Alberdi}, {Osorio}, {G{\'o}mez}, {de Gregorio-Monsalvo},
  {P{\'e}rez-Torres}, {Anglada-Escud{\'e}}, {Berdi{\~n}as}, {Jenkins},
  {Jimenez-Serra}, {Lara}, {L{\'o}pez-Gonz{\'a}lez}, {L{\'o}pez-Puertas},
  {Morales}, {Ribas}, {Richards}, {Rodr{\'{\i}}guez-L{\'o}pez}, \&
  {Rodriguez}}]{Anglada:2018}
{Anglada}, G., {Amado}, P.~J., {Ortiz}, J.~L., {et~al.} 2017, \apjl, 850, L6

\bibitem[{{Babich} {et~al.}(2007){Babich}, {Blake}, \&
  {Steinhardt}}]{Babich:2007}
{Babich}, D., {Blake}, C.~H., \& {Steinhardt}, C.~L. 2007, \apj, 669, 1406

\bibitem[{{Babich} \& {Loeb}(2009)}]{Babich:2009}
{Babich}, D., \& {Loeb}, A. 2009, \na, 14, 166

\bibitem[{{Bauer} {et~al.}(2017){Bauer}, {Grav}, {Fern{\'a}ndez}, {Mainzer},
  {Kramer}, {Masiero}, {Spahr}, {Nugent}, {Stevenson}, {Meech}, {Cutri},
  {Lisse}, {Walker}, {Dailey}, {Rosser}, {Krings}, {Ruecker}, {Wright}, \& {the
  NEOWISE Team}}]{Bauer:2017}
{Bauer}, J.~M., {Grav}, T., {Fern{\'a}ndez}, Y.~R., {et~al.} 2017, \aj, 154, 53

\bibitem[{{Benson} {et~al.}(2014){Benson}, {Ade}, {Ahmed}, {Allen}, {Arnold},
  {Austermann}, {Bender}, {Bleem}, {Carlstrom}, {Chang}, {Cho}, {Cliche},
  {Crawford}, {Cukierman}, {de Haan}, {Dobbs}, {Dutcher}, {Everett}, {Gilbert},
  {Halverson}, {Hanson}, {Harrington}, {Hattori}, {Henning}, {Hilton},
  {Holder}, {Holzapfel}, {Irwin}, {Keisler}, {Knox}, {Kubik}, {Kuo}, {Lee},
  {Leitch}, {Li}, {McDonald}, {Meyer}, {Montgomery}, {Myers}, {Natoli},
  {Nguyen}, {Novosad}, {Padin}, {Pan}, {Pearson}, {Reichardt}, {Ruhl},
  {Saliwanchik}, {Simard}, {Smecher}, {Sayre}, {Shirokoff}, {Stark}, {Story},
  {Suzuki}, {Thompson}, {Tucker}, {Vanderlinde}, {Vieira}, {Vikhlinin}, {Wang},
  {Yefremenko}, \& {Yoon}}]{Benson:2014}
{Benson}, B.~A., {Ade}, P.~A.~R., {Ahmed}, Z., {et~al.} 2014, in \procspie,
  Vol. 9153, Millimeter, Submillimeter, and Far-Infrared Detectors and
  Instrumentation for Astronomy VII, 91531P

\bibitem[{{Bernstein} {et~al.}(2004){Bernstein}, {Trilling}, {Allen}, {Brown},
  {Holman}, \& {Malhotra}}]{Bernstein:2004}
{Bernstein}, G.~M., {Trilling}, D.~E., {Allen}, R.~L., {et~al.} 2004, \aj, 128,
  1364

\bibitem[{{Boller} {et~al.}(2016){Boller}, {Freyberg}, {Tr{\"u}mper}, {Haberl},
  {Voges}, \& {Nandra}}]{Boller:2016}
{Boller}, T., {Freyberg}, M.~J., {Tr{\"u}mper}, J., {et~al.} 2016, \aap, 588,
  A103

\bibitem[{{Brown} \& {Butler}(2017)}]{Brown:2017}
{Brown}, M.~E., \& {Butler}, B.~J. 2017, \aj, 154, 19

\bibitem[{{Brown} {et~al.}(2004){Brown}, {Trujillo}, \&
  {Rabinowitz}}]{Brown:2004}
{Brown}, M.~E., {Trujillo}, C., \& {Rabinowitz}, D. 2004, \apj, 617, 645

\bibitem[{Carlstrom {et~al.}(2011)Carlstrom, Ade, Aird, Benson, Bleem, Busetti,
  Chang, Chauvin, Cho, Crawford, Crites, Dobbs, Halverson, Heimsath, Holzapfel,
  Hrubes, Joy, Keisler, Lanting, Lee, Leitch, Leong, Lu, Lueker, Luong-Van,
  McMahon, Mehl, Meyer, Mohr, Montroy, Padin, Plagge, Pryke, Ruhl, Schaffer,
  Schwan, Shirokoff, Spieler, Staniszewski, Stark, Tucker, Vanderlinde, Vieira,
  \& Williamson}]{Carlstrom:2011}
Carlstrom, J.~E., Ade, P. A.~R., Aird, K.~A., {et~al.} 2011, Publications of
  the Astronomical Society of the Pacific, 123, 568.
\newblock \url{http://stacks.iop.org/1538-3873/123/i=903/a=568}

\bibitem[{{Condon} {et~al.}(1998){Condon}, {Cotton}, {Greisen}, {Yin},
  {Perley}, {Taylor}, \& {Broderick}}]{Condon:1998}
{Condon}, J.~J., {Cotton}, W.~D., {Greisen}, E.~W., {et~al.} 1998, \aj, 115,
  1693

\bibitem[{{Cowan} {et~al.}(2016){Cowan}, {Holder}, \& {Kaib}}]{Cowan:2016}
{Cowan}, N.~B., {Holder}, G., \& {Kaib}, N.~A. 2016, \apjl, 822, L2

\bibitem[{{David} \& {Hillenbrand}(2015)}]{David:2015}
{David}, T.~J., \& {Hillenbrand}, L.~A. 2015, \apj, 804, 146

\bibitem[{{Dicker} {et~al.}(2014){Dicker}, {Ade}, {Aguirre}, {Brevik}, {Cho},
  {Datta}, {Devlin}, {Dober}, {Egan}, {Ford}, {Ford}, {Hilton}, {Irwin},
  {Mason}, {Marganian}, {Mello}, {McMahon}, {Mroczkowski}, {Rosenman},
  {Tucker}, {Vale}, {White}, {Whitehead}, \& {Young}}]{Mustang2}
{Dicker}, S.~R., {Ade}, P.~A.~R., {Aguirre}, J., {et~al.} 2014, Journal of Low
  Temperature Physics, 176, 808

\bibitem[{{Dober} {et~al.}(2014){Dober}, {Ade}, {Ashton}, {Angil{\`e}},
  {Beall}, {Becker}, {Bradford}, {Che}, {Cho}, {Devlin}, {Fissel}, {Fukui},
  {Galitzki}, {Gao}, {Groppi}, {Hillbrand}, {Hilton}, {Hubmayr}, {Irwin},
  {Klein}, {Van Lanen}, {Li}, {Li}, {Lourie}, {Mani}, {Martin}, {Mauskopf},
  {Nakamura}, {Novak}, {Pappas}, {Pascale}, {Santos}, {Savini}, {Scott},
  {Stanchfield}, {Ullom}, {Underhill}, {Vissers}, \&
  {Ward-Thompson}}]{Blastpol}
{Dober}, B.~J., {Ade}, P.~A.~R., {Ashton}, P., {et~al.} 2014, in \procspie,
  Vol. 9153, Millimeter, Submillimeter, and Far-Infrared Detectors and
  Instrumentation for Astronomy VII, 91530H

\bibitem[{{Dones} {et~al.}(2015){Dones}, {Brasser}, {Kaib}, \&
  {Rickman}}]{Dones:2015}
{Dones}, L., {Brasser}, R., {Kaib}, N., \& {Rickman}, H. 2015, \ssr, 197, 191

\bibitem[{{Dones} {et~al.}(2004){Dones}, {Weissman}, {Levison}, \&
  {Duncan}}]{Dones:2004}
{Dones}, L., {Weissman}, P.~R., {Levison}, H.~F., \& {Duncan}, M.~J. 2004, in
  Astronomical Society of the Pacific Conference Series, Vol. 323, Star
  Formation in the Interstellar Medium: In Honor of David Hollenbach, ed.
  D.~{Johnstone}, F.~C. {Adams}, D.~N.~C. {Lin}, D.~A. {Neufeeld}, \& E.~C.
  {Ostriker}, 371

\bibitem[{{Duncan} {et~al.}(1987){Duncan}, {Quinn}, \&
  {Tremaine}}]{Duncan:1987}
{Duncan}, M., {Quinn}, T., \& {Tremaine}, S. 1987, \aj, 94, 1330

\bibitem[{{Dwek} {et~al.}(1980){Dwek}, {Sellgren}, {Soifer}, \&
  {Werner}}]{Dwek:1980}
{Dwek}, E., {Sellgren}, K., {Soifer}, B.~T., \& {Werner}, M.~W. 1980, \apj,
  238, 140

\bibitem[{{Eker} {et~al.}(2018){Eker}, {Bakis}, {Bilir}, {Soydugan}, {Steer},
  {Soydugan}, {Bakis}, {Alicavus}, {Aslan}, \& {Alpsoy}}]{Eker:2018}
{Eker}, Z., {Bakis}, V., {Bilir}, S., {et~al.} 2018, ArXiv e-prints,
  arXiv:1807.02568

\bibitem[{{Gaia Collaboration} {et~al.}(2018){Gaia Collaboration}, {Brown},
  {Vallenari}, {Prusti}, {de Bruijne}, {Babusiaux}, \&
  {Bailer-Jones}}]{Gaia:2018}
{Gaia Collaboration}, {Brown}, A.~G.~A., {Vallenari}, A., {et~al.} 2018, ArXiv
  e-prints, arXiv:1804.09365

\bibitem[{{G{\"u}del}(2002)}]{Gudel:2002}
{G{\"u}del}, M. 2002, \araa, 40, 217

\bibitem[{{Habing} {et~al.}(1999){Habing}, {Dominik}, {Jourdain de Muizon},
  {Kessler}, {Laureijs}, {Leech}, {Metcalfe}, {Salama}, {Siebenmorgen}, \&
  {Trams}}]{Habing:1999}
{Habing}, H.~J., {Dominik}, C., {Jourdain de Muizon}, M., {et~al.} 1999, \nat,
  401, 456

\bibitem[{{Henderson} {et~al.}(2016){Henderson}, {Allison}, {Austermann},
  {Baildon}, {Battaglia}, {Beall}, {Becker}, {De Bernardis}, {Bond},
  {Calabrese}, {Choi}, {Coughlin}, {Crowley}, {Datta}, {Devlin}, {Duff},
  {Dunkley}, {D{\"u}nner}, {van Engelen}, {Gallardo}, {Grace}, {Hasselfield},
  {Hills}, {Hilton}, {Hincks}, {Hlo{\^z}ek}, {Ho}, {Hubmayr}, {Huffenberger},
  {Hughes}, {Irwin}, {Koopman}, {Kosowsky}, {Li}, {McMahon}, {Munson}, {Nati},
  {Newburgh}, {Niemack}, {Niraula}, {Page}, {Pappas}, {Salatino}, {Schillaci},
  {Schmitt}, {Sehgal}, {Sherwin}, {Sievers}, {Simon}, {Spergel}, {Staggs},
  {Stevens}, {Thornton}, {Van Lanen}, {Vavagiakis}, {Ward}, \&
  {Wollack}}]{Henderson:2016}
{Henderson}, S.~W., {Allison}, R., {Austermann}, J., {et~al.} 2016, Journal of
  Low Temperature Physics, 184, 772

\bibitem[{{Holler} {et~al.}(2017){Holler}, {Milam}, {Bauer}, {Alcock},
  {Bannister}, {Bjoraker}, {Bodewits}, {Bosh}, {Buie}, {Farnham},
  {Haghighipour}, {Hardersen}, {Harris}, {Hirata}, {Hsieh}, {Kelley}, {Knight},
  {Kramer}, {Longobardo}, {Nixon}, {Palomba}, {Protopapa}, {Quick},
  {Ragozzine}, {Reddy}, {Rhodes}, {Rivkin}, {Sarid}, {Sickafoose}, {Simon},
  {Thomas}, {Trilling}, \& {West}}]{Holler:2017}
{Holler}, B.~J., {Milam}, S.~N., {Bauer}, J.~M., {et~al.} 2017, ArXiv e-prints,
  arXiv:1709.02763

\bibitem[{{Howe} \& {Rafikov}(2014)}]{Howe:2014}
{Howe}, A.~R., \& {Rafikov}, R.~R. 2014, \apj, 781, 52

\bibitem[{{Hughes} {et~al.}(2018){Hughes}, {Duchene}, \&
  {Matthews}}]{Hughes:2018}
{Hughes}, A.~M., {Duchene}, G., \& {Matthews}, B. 2018, ArXiv e-prints,
  arXiv:1802.04313

\bibitem[{{Ichikawa} \& {Fukugita}(2011)}]{Ichikawa:2011}
{Ichikawa}, K., \& {Fukugita}, M. 2011, \apj, 736, 122

\bibitem[{{K{\'o}sp{\'a}l} {et~al.}(2013){K{\'o}sp{\'a}l}, {Mo{\'o}r},
  {Juh{\'a}sz}, {{\'A}brah{\'a}m}, {Apai}, {Csengeri}, {Grady}, {Henning},
  {Hughes}, {Kiss}, {Pascucci}, \& {Schmalzl}}]{Kospal:2013}
{K{\'o}sp{\'a}l}, {\'A}., {Mo{\'o}r}, A., {Juh{\'a}sz}, A., {et~al.} 2013,
  \apj, 776, 77

\bibitem[{{Lehner} {et~al.}(2016){Lehner}, {Wang}, {Reyes-Ruiz}, {Alcock},
  {Castro}, {Chen}, {Chu}, {Cook}, {Figueroa}, {Geary}, {Huang}, {Kim},
  {Norton}, {Szentgyorgyi}, {Yen}, \& {Zhang}}]{Lehner:2016}
{Lehner}, M.~J., {Wang}, S.-Y., {Reyes-Ruiz}, M., {et~al.} 2016, in \procspie,
  Vol. 9906, Ground-based and Airborne Telescopes VI, 99065M

\bibitem[{{Lenz} {et~al.}(2017){Lenz}, {Hensley}, \& {Dor{\'e}}}]{Lenz:2017}
{Lenz}, D., {Hensley}, B.~S., \& {Dor{\'e}}, O. 2017, \apj, 846, 38

\bibitem[{{LSST Science Collaboration} {et~al.}(2009){LSST Science
  Collaboration}, {Abell}, {Allison}, {Anderson}, {Andrew}, {Angel}, {Armus},
  {Arnett}, {Asztalos}, {Axelrod}, \& et~al.}]{LSSTsciencebook}
{LSST Science Collaboration}, {Abell}, P.~A., {Allison}, J., {et~al.} 2009,
  ArXiv e-prints, arXiv:0912.0201

\bibitem[{{MacGregor} {et~al.}(2017){MacGregor}, {Matr{\`a}}, {Kalas},
  {Wilner}, {Pan}, {Kennedy}, {Wyatt}, {Duchene}, {Hughes}, {Rieke}, {Clampin},
  {Fitzgerald}, {Graham}, {Holland}, {Pani{\'c}}, {Shannon}, \&
  {Su}}]{MacGregor:2017}
{MacGregor}, M.~A., {Matr{\`a}}, L., {Kalas}, P., {et~al.} 2017, \apj, 842, 8

\bibitem[{{Mathys}(2017)}]{Mathys:2017}
{Mathys}, G. 2017, \aap, 601, A14

\bibitem[{{McDonald} {et~al.}(2017){McDonald}, {Zijlstra}, \&
  {Watson}}]{McDonald:2017}
{McDonald}, I., {Zijlstra}, A.~A., \& {Watson}, R.~A. 2017, \mnras, 471, 770

\bibitem[{{Morales} {et~al.}(2016){Morales}, {Bryden}, {Werner}, \&
  {Stapelfeldt}}]{Morales:2016}
{Morales}, F.~Y., {Bryden}, G., {Werner}, M.~W., \& {Stapelfeldt}, K.~R. 2016,
  \apj, 831, 97

\bibitem[{{Norberg} {et~al.}(2009){Norberg}, {Baugh}, {Gazta{\~n}aga}, \&
  {Croton}}]{Norberg:2009}
{Norberg}, P., {Baugh}, C.~M., {Gazta{\~n}aga}, E., \& {Croton}, D.~J. 2009,
  \mnras, 396, 19

\bibitem[{{Ofek} \& {Nakar}(2010)}]{Ofek:2010}
{Ofek}, E.~O., \& {Nakar}, E. 2010, \apjl, 711, L7

\bibitem[{{Oort}(1950)}]{Oort:1950}
{Oort}, J.~H. 1950, \bain, 11, 91

\bibitem[{{P{\'a}l} {et~al.}(2012){P{\'a}l}, {Kiss}, {M{\"u}ller},
  {Santos-Sanz}, {Vilenius}, {Szalai}, {Mommert}, {Lellouch}, {Rengel},
  {Hartogh}, {Protopapa}, {Stansberry}, {Ortiz}, {Duffard}, {Thirouin},
  {Henry}, \& {Delsanti}}]{Pal:2012}
{P{\'a}l}, A., {Kiss}, C., {M{\"u}ller}, T.~G., {et~al.} 2012, \aap, 541, L6

\bibitem[{{Pan} \& {Sari}(2005)}]{Pan:2005}
{Pan}, M., \& {Sari}, R. 2005, \icarus, 173, 342

\bibitem[{{Planck Collaboration} {et~al.}(2014){Planck Collaboration}, {Ade},
  {Aghanim}, {Armitage-Caplan}, {Arnaud}, {Ashdown}, {Atrio-Barandela},
  {Aumont}, {Baccigalupi}, {Banday}, \& et~al.}]{Planck:conversion}
{Planck Collaboration}, {Ade}, P.~A.~R., {Aghanim}, N., {et~al.} 2014, \aap,
  571, A9

\bibitem[{{Planck Collaboration} {et~al.}(2016){Planck Collaboration}, {Adam},
  {Ade}, {Aghanim}, {Arnaud}, {Ashdown}, {Aumont}, {Baccigalupi}, {Banday},
  {Barreiro}, \& et~al.}]{Planck:HFI}
{Planck Collaboration}, {Adam}, R., {Ade}, P.~A.~R., {et~al.} 2016, \aap, 594,
  A8

\bibitem[{{Planck Collaboration} {et~al.}(2018){Planck Collaboration},
  {Akrami}, {Arroja}, {Ashdown}, {Aumont}, {Baccigalupi}, {Ballardini},
  {Banday}, {Barreiro}, {Bartolo}, {Basak}, {Battye}, {Benabed}, {Bernard},
  {Bersanelli}, {Bielewicz}, {Bock}, {Bond}, {Borrill}, {Bouchet}, {Boulanger},
  {Bucher}, {Burigana}, {Butler}, {Calabrese}, {Cardoso}, {Carron},
  {Casaponsa}, {Challinor}, {Chiang}, {Colombo}, {Combet}, {Contreras},
  {Crill}, {Cuttaia}, {de Bernardis}, {de Zotti}, {Delabrouille}, {Delouis},
  {D{\'e}sert}, {Di Valentino}, {Dickinson}, {Diego}, {Donzelli}, {Dor{\'e}},
  {Douspis}, {Ducout}, {Dupac}, {Efstathiou}, {Elsner}, {En{\ss}lin},
  {Eriksen}, {Falgarone}, {Fantaye}, {Fergusson}, {Fernandez-Cobos}, {Finelli},
  {Forastieri}, {Frailis}, {Franceschi}, {Frolov}, {Galeotta}, {Galli},
  {Ganga}, {G{\'e}nova-Santos}, {Gerbino}, {Ghosh}, {Gonz{\'a}lez-Nuevo},
  {G{\'o}rski}, {Gratton}, {Gruppuso}, {Gudmundsson}, {Hamann}, {Handley},
  {Hansen}, {Helou}, {Herranz}, {Hivon}, {Huang}, {Jaffe}, {Jones}, {Karakci},
  {Keih{\"a}nen}, {Keskitalo}, {Kiiveri}, {Kim}, {Kisner}, {Knox},
  {Krachmalnicoff}, {Kunz}, {Kurki-Suonio}, {Lagache}, {Lamarre}, {Langer},
  {Lasenby}, {Lattanzi}, {Lawrence}, {Le Jeune}, {Leahy}, {Lesgourgues},
  {Levrier}, {Lewis}, {Liguori}, {Lilje}, {Lilley}, {Lindholm},
  {L{\'o}pez-Caniego}, {Lubin}, {Ma}, {Mac{\'{\i}}as-P{\'e}rez}, {Maggio},
  {Maino}, {Mandolesi}, {Mangilli}, {Marcos-Caballero}, {Maris}, {Martin},
  {Mart{\'{\i}}nez-Gonz{\'a}lez}, {Matarrese}, {Mauri}, {McEwen}, {Meerburg},
  {Meinhold}, {Melchiorri}, {Mennella}, {Migliaccio}, {Millea}, {Mitra},
  {Miville-Desch{\^e}nes}, {Molinari}, {Moneti}, {Montier}, {Morgante}, {Moss},
  {Mottet}, {M{\"u}nchmeyer}, {Natoli}, {N{\o}rgaard-Nielsen}, {Oxborrow},
  {Pagano}, {Paoletti}, {Partridge}, {Patanchon}, {Pearson}, {Peel}, {Peiris},
  {Perrotta}, {Pettorino}, {Piacentini}, {Polastri}, {Polenta}, {Puget},
  {Rachen}, {Reinecke}, {Remazeilles}, {Renzi}, {Rocha}, {Rosset}, {Roudier},
  {Rubi{\~n}o-Mart{\'{\i}}n}, {Ruiz-Granados}, {Salvati}, {Sandri},
  {Savelainen}, {Scott}, {Shellard}, {Shiraishi}, {Sirignano}, {Sirri},
  {Spencer}, {Sunyaev}, {Suur-Uski}, {Tauber}, {Tavagnacco}, {Tenti},
  {Terenzi}, {Toffolatti}, {Tomasi}, {Trombetti}, {Valiviita}, {Van Tent},
  {Vibert}, {Vielva}, {Villa}, {Vittorio}, {Wandelt}, {Wehus}, {White},
  {White}, {Zacchei}, \& {Zonca}}]{Planck:2018}
{Planck Collaboration}, {Akrami}, Y., {Arroja}, F., {et~al.} 2018, ArXiv
  e-prints, arXiv:1807.06205

\bibitem[{{Rappaport} {et~al.}(2018){Rappaport}, {Vanderburg}, {Jacobs},
  {LaCourse}, {Jenkins}, {Kraus}, {Rizzuto}, {Latham}, {Bieryla}, {Lazarevic},
  \& {Schmitt}}]{Rappaport:2018}
{Rappaport}, S., {Vanderburg}, A., {Jacobs}, T., {et~al.} 2018, \mnras, 474,
  1453

\bibitem[{{Roberge} {et~al.}(2013){Roberge}, {Kamp}, {Montesinos}, {Dent},
  {Meeus}, {Donaldson}, {Olofsson}, {Mo{\'o}r}, {Augereau}, {Howard}, {Eiroa},
  {Thi}, {Ardila}, {Sandell}, \& {Woitke}}]{Roberge:2013}
{Roberge}, A., {Kamp}, I., {Montesinos}, B., {et~al.} 2013, \apj, 771, 69

\bibitem[{{Saunders} {et~al.}(2000){Saunders}, {Sutherland}, {Maddox},
  {Keeble}, {Oliver}, {Rowan-Robinson}, {McMahon}, {Efstathiou}, {Tadros},
  {White}, {Frenk}, {Carrami{\~n}ana}, \& {Hawkins}}]{Saunders:2000}
{Saunders}, W., {Sutherland}, W.~J., {Maddox}, S.~J., {et~al.} 2000, \mnras,
  317, 55

\bibitem[{{Shannon} {et~al.}(2015){Shannon}, {Jackson}, {Veras}, \&
  {Wyatt}}]{Shannon:2015}
{Shannon}, A., {Jackson}, A.~P., {Veras}, D., \& {Wyatt}, M. 2015, \mnras, 446,
  2059

\bibitem[{{Sibthorpe} {et~al.}(2010){Sibthorpe}, {Vandenbussche}, {Greaves},
  {Pantin}, {Olofsson}, {Acke}, {Barlow}, {Blommaert}, {Bouwman}, {Brandeker},
  {Cohen}, {De Meester}, {Dent}, {di Francesco}, {Dominik}, {Fridlund}, {Gear},
  {Glauser}, {Gomez}, {Hargrave}, {Harvey}, {Henning}, {Heras}, {Hogerheijde},
  {Holland}, {Ivison}, {Leeks}, {Lim}, {Liseau}, {Matthews}, {Naylor},
  {Pilbratt}, {Polehampton}, {Regibo}, {Royer}, {Sicilia-Aguilar}, {Swinyard},
  {Waelkens}, {Walker}, \& {Wesson}}]{Sibthorpe:2010}
{Sibthorpe}, B., {Vandenbussche}, B., {Greaves}, J.~S., {et~al.} 2010, \aap,
  518, L130

\bibitem[{{Simon} {et~al.}(1995){Simon}, {Drake}, \& {Kim}}]{Simon:1995}
{Simon}, T., {Drake}, S.~A., \& {Kim}, P.~D. 1995, \pasp, 107, 1034

\bibitem[{{Stern} {et~al.}(1991){Stern}, {Stocke}, \& {Weissman}}]{Stern:1991}
{Stern}, S.~A., {Stocke}, J., \& {Weissman}, P.~R. 1991, \icarus, 91, 65

\bibitem[{{Stone} {et~al.}(2015){Stone}, {Metzger}, \& {Loeb}}]{Stone:2015}
{Stone}, N., {Metzger}, B.~D., \& {Loeb}, A. 2015, \mnras, 448, 188

\bibitem[{{Swetz} {et~al.}(2011){Swetz}, {Ade}, {Amiri}, {Appel},
  {Battistelli}, {Burger}, {Chervenak}, {Devlin}, {Dicker}, {Doriese},
  {D{\"u}nner}, {Essinger-Hileman}, {Fisher}, {Fowler}, {Halpern},
  {Hasselfield}, {Hilton}, {Hincks}, {Irwin}, {Jarosik}, {Kaul}, {Klein},
  {Lau}, {Limon}, {Marriage}, {Marsden}, {Martocci}, {Mauskopf}, {Moseley},
  {Netterfield}, {Niemack}, {Nolta}, {Page}, {Parker}, {Staggs}, {Stryzak},
  {Switzer}, {Thornton}, {Tucker}, {Wollack}, \& {Zhao}}]{Swetz2011}
{Swetz}, D.~S., {Ade}, P.~A.~R., {Amiri}, M., {et~al.} 2011, \apjs, 194, 41

\bibitem[{{The Simons Observatory Collaboration} {et~al.}(2018){The Simons
  Observatory Collaboration}, {Ade}, {Aguirre}, {Ahmed}, {Aiola}, {Ali},
  {Alonso}, {Alvarez}, {Arnold}, {Ashton}, \& et~al.}]{SOforecast}
{The Simons Observatory Collaboration}, {Ade}, P., {Aguirre}, J., {et~al.}
  2018, ArXiv e-prints, arXiv:1808.07445

\bibitem[{{Trilling} {et~al.}(2018){Trilling}, {Bellm}, \&
  {Malhotra}}]{Trilling:2018}
{Trilling}, D.~E., {Bellm}, E.~C., \& {Malhotra}, R. 2018, \aj, 155, 243

\bibitem[{{Ventura} {et~al.}(2018){Ventura}, {Karakas}, {Dell'Agli},
  {Garc{\'{\i}}a-Hern{\'a}ndez}, \& {Guzman-Ramirez}}]{Ventura:2018}
{Ventura}, P., {Karakas}, A., {Dell'Agli}, F., {Garc{\'{\i}}a-Hern{\'a}ndez},
  D.~A., \& {Guzman-Ramirez}, L. 2018, \mnras, 475, 2282

\bibitem[{{Welsh} \& {Montgomery}(2016)}]{Welsh:2016}
{Welsh}, B.~Y., \& {Montgomery}, S. 2016, \pasp, 128, 064201

\bibitem[{{White} {et~al.}(2018){White}, {Aufdenberg}, {Boley}, {Hauschildt},
  {Hughes}, {Matthews}, \& {Wilner}}]{White:2018}
{White}, J.~A., {Aufdenberg}, J., {Boley}, A.~C., {et~al.} 2018, \apj, 859, 102

\bibitem[{{Wootten} \& {Thompson}(2009)}]{ALMA}
{Wootten}, A., \& {Thompson}, A.~R. 2009, IEEE Proceedings, 97, 1463

\end{thebibliography}

\appendix

\section{Covariance matrices of stacked 545 and 857 GHz measurements around {\it Gaia} stars}

\begin{figure}
\centering
\includegraphics[scale=0.5]{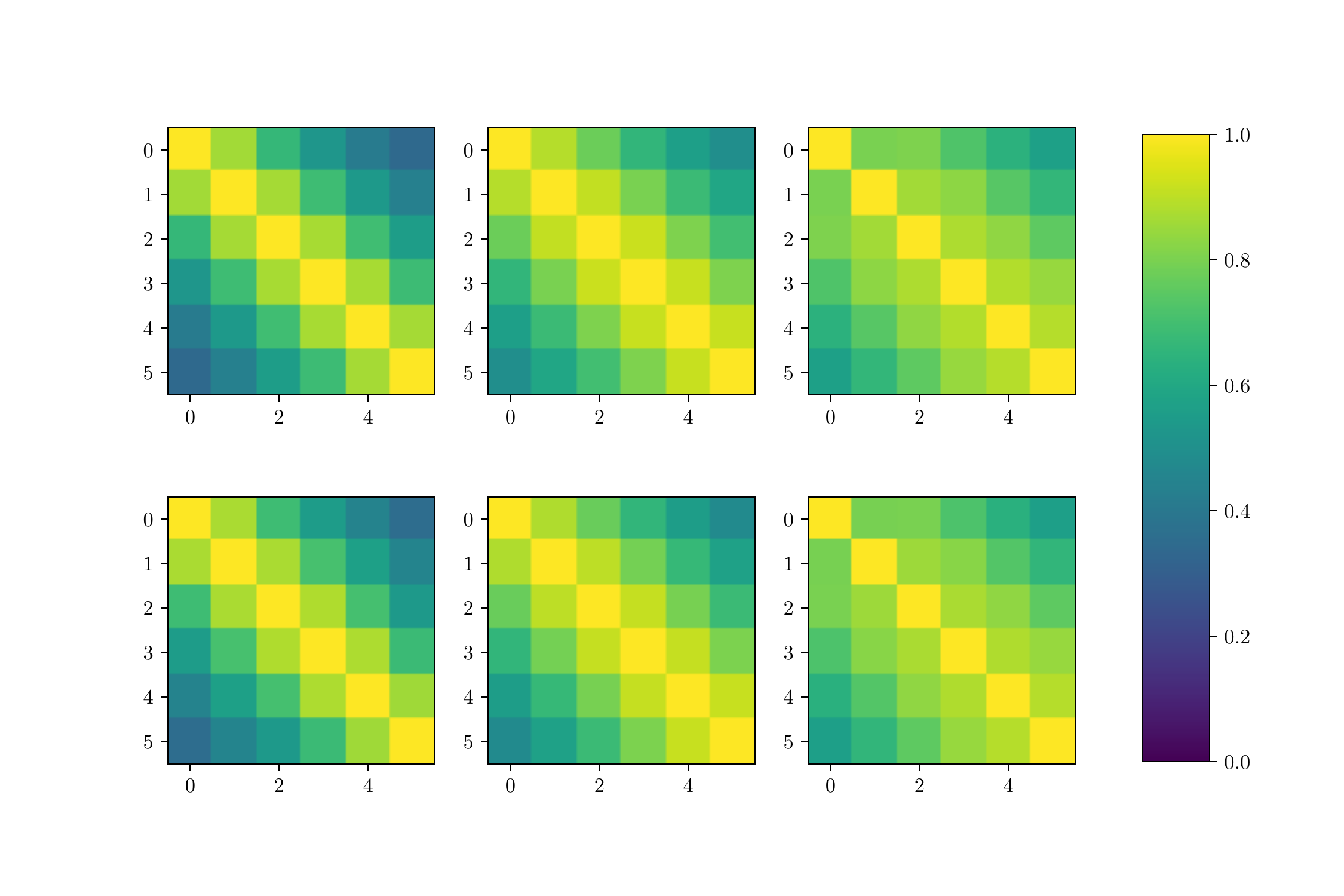}
\caption{\label{fig:correlation_matrix_545} Correlation matrices corresponding to the 545 GHz  measurements plotted in Fig.~\ref{fig:measurement}.}
\end{figure}

\begin{figure}
\centering
\includegraphics[scale=0.5]{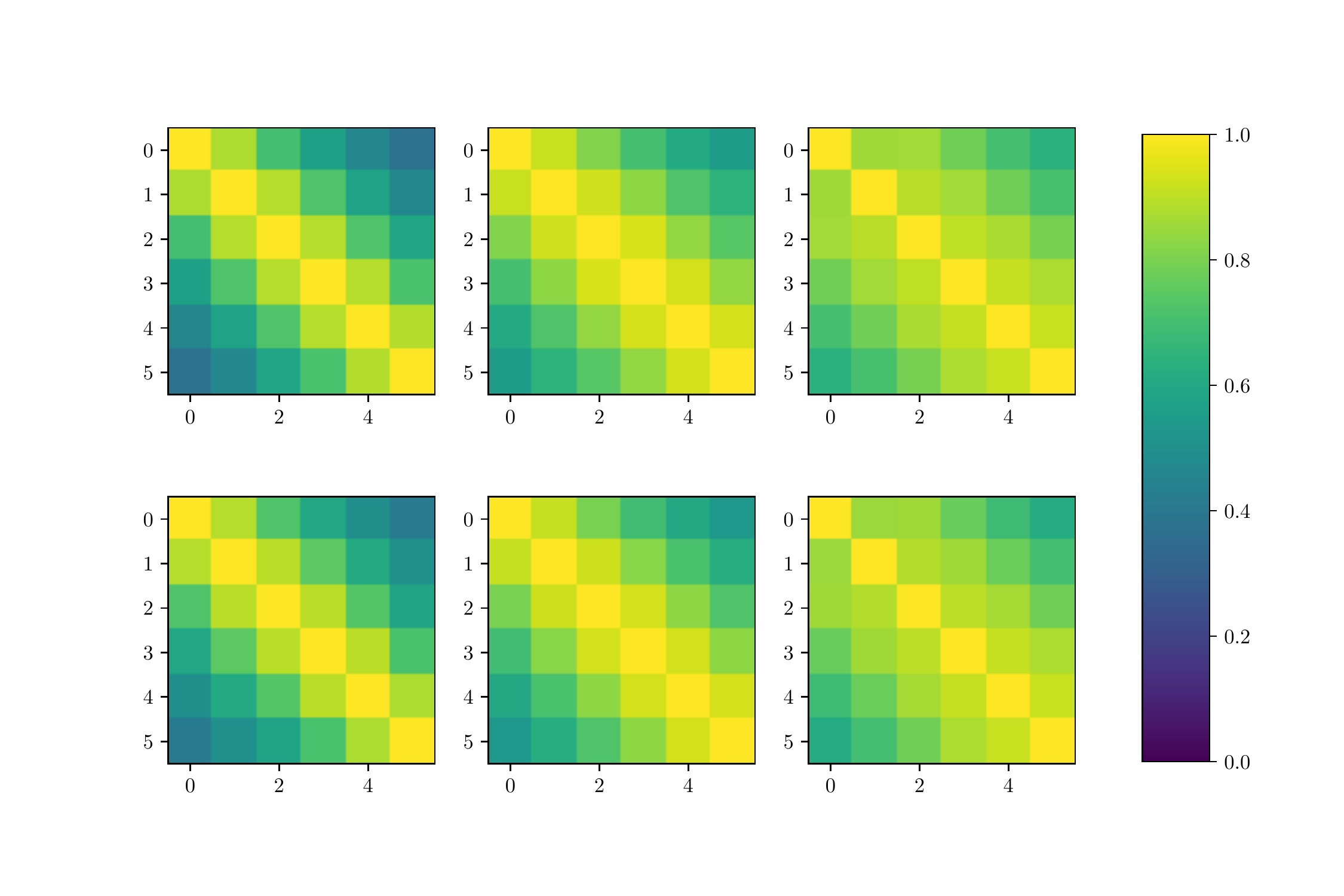}
\caption{\label{fig:correlation_matrix_857} Correlation matrices corresponding to the 857 GHz  measurements plotted in Fig.~\ref{fig:measurement}.}
\end{figure}

Figs.~\ref{fig:correlation_matrix_545} and ~\ref{fig:correlation_matrix_857}  show the correlation matrices (i.e. $C_{ij}/\sqrt{C_{ii}C_{jj}}$, where $C_{ij}$ is an element of the covariance matrix) for the 545 and 857 GHz measurements plotted in Fig.~\ref{fig:measurement}, respectively.  The ordering of the panels is the same as in Fig.~\ref{fig:measurement}: moving top to bottom corresponds to stacking around fainter stars, while moving left to right corresponds to stacking around more distant stars.  More distant stars (right panels) correspond to smaller angular scales on the sky, and therefore result in more covariant measurements due to the effects of the {\it Planck} beam.

\end{document}